\documentclass[twocolumn,pra,superscriptaddress,showpacs,aps,floatfix]{revtex4-2}
\usepackage{color}
\usepackage{amsmath}
\usepackage{amssymb}
\usepackage{txfonts}
\usepackage{graphicx}
\usepackage{epstopdf}
\usepackage[unicode]{hyperref}
\hypersetup{pdftitle={2D SOC BEC},
 pdfauthor={},
 pdfsubject={2D SOC BEC},
  colorlinks=true,
    linkcolor=red,
    citecolor=magenta,
    filecolor=black,
    urlcolor=blue}


\begin{document}

\title{BAU production in the $S_N$-breaking Standard Model}

\author{Chilong Lin}
\email{lingo@mail.nmns.edu.tw}

\affiliation{National Museum of Natural Science, 1st, Guan Chien RD., Taichung, 40453 Taiwan, ROC}

\date{Version of \today. }

\begin{abstract}

CP violation (CPV) and the baryon asymmetry of the universe (BAU) are two of the most significant unresolved problems in physics.
This article presents further research on the CPV problem in the standard model (SM) with thirty-two candidate sets of the ten "natural" parameters that exhibit the same CKM performance.
These parameters are considered "natural" because they consist solely of the Yukawa couplings and the vacuum expectation value (VEV) of the unique Higgs doublet in the SM.
We then investigate the CPV problem and the BAU problem using the Jarlskog measure of CPV,
  $\Delta_{CP}=J(m^2_t -m^2_c)(m^2_t -m^2_u)(m^2_c -m^2_u)(m^2_b -m^2_s)(m^2_b -m^2_d)(m^2_s -m^2_d)/ T^{12}$,
   given that CP symmetry is violated following the breakdowns of $S_N$ symmetries.
Subsequently, we perform numerical tests in a simplified scenario where eight of the ten parameters are fixed, and the remaining two parameters are allowed to vary from their $S_2$-symmetric values ${\bf (x,~y)}=(-1,~1)$ to their current values in all thirty-two parameter sets.
To estimate the enhancement of CPV in such processes,
 we propose a ratio $R_{\Delta} \equiv \Delta_{CP} / \Delta^{(0)}_{CP}$ between the running $\Delta_{CP}$ and its current value,
 denoted by $\Delta^{(0)}_{CP}$, which is approximately $10^{-20}$.
In all thirty-two cases, the three-dimensional (3D) plots of $R_{\Delta}$ exhibit many regions that stick out of the $R_{\Delta}=10^{10}$ plane,
 especially in regions very close to ${\bf (x,~y)}=(-1,~1)$.
These results demonstrate that the $S_N$-breaking Standard Model is already sufficient to violate CP symmetry explicitly and generate a significant amount of BAU.
Furthermore, it solves existing problems without creating new ones.

\end{abstract}
\maketitle


\section{Introduction}

The problem of how CP symmetry was violated in electroweak interactions is a long-unsolved problem since it was discovered in 1964 \cite{Christenson1964}.
In the Standard Model (SM) of electroweak interactions, CP violation (CPV) comes solely from a complex phase in the Cabibbo-Kobayashi-Maskawa (CKM) matrix  \cite{Cabibbo1963, KM1973}.
However, the CKM matrix is a product of two unitary transformation matrices that diagonalize the mass matrices of up- and down-type quarks, respectively.
Naturally, the key to igniting CPV is concealed in the quark mass matrices. \\

In several of our previous researches, we found that in SM and in its extension with one extra Higgs doublet, the two-Higgs doublet model (2HDM),
$S_3$ symmetries among three fermion generations make the CP symmetry always conserved \cite{Lee1986a, Lee1986b, Lin1988, Lee1990, Lin1994}.
However, if the $S_3$ symmetries were broken down to various residual $S_2$ symmetries,
a complex phase appears explicitly in the CKM matrix,
 which means that CP symmetry was violated following the breakdowns of $S_3$ symmetries \cite{Lin2020a, Lin2021}.
The Jarlskog invariant $J\sim 0.171$ thus derived is about 5,700 times that of ${\it J^{(0)}}=(3.00^{+0.15}_{-0.09}) \times 10^{-5}$ given by current experiments \cite{Zyla2020}. \\

In \cite{Lin2019, Lin2021}, the $S_2$ symmetries were further broken down to completely nothing and the derived CKM matrix contains only four parameters,
two from up-type quarks and two from down-type quarks.
These four parameters are natural parameters since they are solely composed of the Yukawa couplings and $v$, the vacuum expectation value (VEV) of the only Higgs doublet in SM.
The CKM  elements thus derived fit experiments to an accuracy of ${\bf O}(\lambda^{1/2})$ or better at tree level. \\

How and when the Baryon Asymmetry of the Universe (BAU) \cite{Ade2014, Ade2016, Bennet2003} was produced and left until now is another long unsolved problem in physics.
In cosmology, BAU is estimated by the excess of baryons over anti-baryons, denoted by ${\it \eta =(n_B -n_{\bar B})/n_{\gamma} }$ or ${\it (n_B -n_{\bar B})/s }$,
where $n_B$, $n_{\bar B}$, $n_{\gamma}$, and $s$ are number densities of baryon, anti-baryon, photon, and entropy density observed in current universe, respectively.
Combining the result of the $\bf Planck$ \cite{Ade2014, Ade2016} with that of the $\bf WMAP$ \cite{Bennet2003}, ${\it \eta}~ \approx 10^{-10}$. \\

However, such a quantity is measured in the standard model of particle physics by the Jarlskog measure of CPV \cite{Jarlskog1985, Tranberg2011},
\begin{eqnarray}
\Delta_{CP} &=& {\rm Im~ Det} [m_u m_u^{\dagger} , m_d m_d^{\dagger} ] /T^{12} \nonumber \\
            &=& {\it J}~ \prod_{\scriptstyle i < j} (m_{u,i}^2 - m_{u,j}^2 ) \prod_{\scriptstyle i < j}(m_{d,i}^2 - m_{d,j}^2 ) / T^{12}.
\end{eqnarray}
Here, $J$ is the Jarlskog invariant, $T\approx$ 100 GeV is the temperature of the electroweak phase transition and $m^2$ are squares of quark masses.
In many other studies, the factor $T^{12}$ has been replaced by $v^{12}$, where $v=$ 246 Gev is the VEV of the SM Higgs doublet.
However, in this article, choosing between $T^{12}$ or $v^{12}$ is not a problem at all,
 since such a factor will be canceled out automatically, as will be shown in section III. \\

Substituting all parameters in $\Delta_{CP}$ with their current empirical values,
$\Delta_{CP}^{(0)}$ is given by $\approx 10^{-20}$ \cite{Shaposhnikov1986, Shaposhnikov1993, Gavela1994, Huet1995, Zyla2020},
which is much smaller than the cosmologically observed ${\it \eta}~ \approx 10^{-10}$.
The discrepancy between these two quantities is what is called the BAU problem  \cite{Ade2014, Ade2016, Bennet2003}.
\\

Considering that the Jarlskog measure of CPV contains a component related to $m^2$ and a CKM-related Jarlskog invariant, given by
\begin{eqnarray}
{\rm Im} [V_{ij} V_{kl} V_{il}^* V_{kj}^* ]={\it J}~ \Sigma_{m,n} \epsilon_{ikm} \epsilon_{jln},
\end{eqnarray}
where $V_{ij}$ are the $i$th column and $j$th row elements of CKM matrix, and $i,~j$ = 1 to 3.
Due to the unitarity of $V_{CKM}$, $J$ must always be smaller than one.
Obviously, increasing $J$ alone is never enough to account for the $10^{10}$ discrepancy.
However, squared quark masses also contribute to $\Delta_{CP}$, and the mass eigenvalues given in our previous research depend on five parameters for each type of quark.
 This means that varying quark masses could be another avenue for further enhancing the strength of $\Delta_{CP}$.
 This is the topic that will be investigated in this article.
\\

In section II, we summarize and supplement our previous research on the CPV problem, which is fundamental material for the following investigations.
In the spirit of "NATURE is simple and elegant", we prefer to study the problem in the SM, if possible, since it does not employ new physics that usually bring in new problems.
In \cite{Lin1988, Lee1990}, we studied a $S_3$-symmetric 2HDM and found that the CP symmetry is conserved.
In \cite{Lin2020a}, we found three less symmetric and more complicated patterns for the mass matrices when $S_3$ symmetries were broken down to residual $S_2$ symmetries, and we noted that the CP symmetry was violated following the breakdowns of $S_3$ symmetries.
However, the breakdowns of $S_N$ symmetries and the violation of CP symmetry are likely two effects of the same cause rather than being causally related. \\

Furthermore, we also found that for every mass matrix pattern obtained in the 2HDM,
there is always a corresponding pattern in the SM if we re-parameterize each element by a complex Yukawa coupling and the VEV of the unique Higgs doublet in the SM.
Thus, it is sensible to study the problems in the SM since it will not bring in extra problems,
such as the flavor-changing-neutral current (FCNC) problem in the 2HDM,
 which has already been solved in \cite{Lin2019}.
In \cite{Lin2020a}, we further simplified the model by replacing the assumed Hermitian $M$ matrices with naturally Hermitian ${\bf M^2}=M \cdot M^{\dagger}$ matrices.
This reduction in assumptions increases the generality of the study.
Moreover, apart from the SM, we only need a very trivial assumption, which is that $\bf M^2_R$ and $\bf M^2_I$ can be simultaneously and respectively diagonalized by the same unitary matrix, to solve the CPV problem.
The benefit of this concept is that it solves problems instead of introducing extra problems, such as the FCNC problem in the 2HDM or the masses and Yukawa couplings of new quarks in the Fourth-generation extension, $etc.$ \\

In section III, we will substitute these materials into $\Delta_{CP}$ and perform some simple tests on it.
As a consequence, sixty-four candidate parameter sets with the same CKM performance are obtained,
 and half of them are further excluded by their predictions of imaginary quark masses.
We will then employ a ratio $R_{\Delta} \equiv \Delta_{CP} ~/~ \Delta^{(0)}_{CP}$ to compare the running $\Delta_{CP}$ with its current value $\Delta^{(0)}_{CP}$.
Since there are ten parameters in $\Delta_{CP}$, it is always possible to have a $R_{\Delta} ~\ge~10^{10}$ by tuning the parameters.
However, this is really too arbitrary to have any physical meanings. \\

Thus, we concentrate our subsequent study on a very simplified case in which
 six parameters are fixed by their current values, and two are fixed by the $S_2$ symmetry of down-type quarks.
We will then let the parameters $\bf x$ and $\bf y$ run from their $S_2$-symmetric values to their current values and see if there are any parameter spaces in which $R_{\Delta} ~\ge~10^{10}$.
As a consequence, all thirty-two cases exhibit many such areas and it is more interesting that
in all their 3D plots, there is always a high $R_{\Delta}$ area around the $S_2$-symmetric point $\bf (x,~ y)$ = (-1, 1),
while $R_{\Delta}$ = 0 at exactly that point.
This indicates that all these cases can provide a very productive environment for BAU when the $S_2$ symmetry of up-type quarks is just beginning to break down.
Thus, an $S_N$-breaking standard model is already enough to solve both the CPV and BAU problems. \\

In section IV, we conclude this article with very brief discussions. \\

\section{CP Violation in the Standard Model }

To solve the CPV and BAU problems, new physical models like extra Higgs doublets, fourth fermion generation, super-symmetry, or many others have been employed to extend the standard model.
However, many of these new physical models have side effects of introducing new problems.
Even worse, some of them introduces more new problems than old ones they solve.
For instance, the 2HDM introduces an FCNC problem but does not solve the CPV problem, though the FCNC in a 2HDM was recently solved in \cite{Lin2019}.
Thus, we prefer to study these problem in the simplest possible way.
If possible, we hope these problems could be solved in the standard model alone without any extensions.
Hereafter, we show that standard model alone is already enough to violate CP symmetry and to produce a large amount of BAU.
Such an SM solves two problems without introducing any new problems. \\

In the standard model, which has three fermion generations, the most general pattern of its $M^q$ matrices is a $3 \times 3$ matrix that contains nine elements and eighteen independent parameters, nine come from real coefficients and nine from imaginary coefficients, if all elements are complex.
The ideal solution to the CPV problem is to diagonalize such $M^q$ matrices analytically and use the $U^q$ matrices thus obtained to produce a complex phase in the CKM matrix, which is by definition a product of two $U^q$ matrices,
\begin{eqnarray}
V_{CKM}=U^u \cdot U^{d\dagger}.
\end{eqnarray}

Although such a diagonalization is theoretically possible, it is too complicated to carry out practically.
Therefore, physicists employed various assumptions to simplify the $M^q$ pattern down to manageable levels so as to carry on their researches.
Such assumptions may include symmetries (like $Z_2$, $S_N$, etc.), Hermiticity, $ad~hoc$ zero elements, and others.
However, such assumptions or constraints always reduce the generality of the models, leading to differences between the obtained solutions and the reality.
The degree of difference between the reality and the predictions depends on the strength of the employed constraints.
In other words, models with fewer assumptions will be closer to reality, and the ideal solution is the one obtained by directly diagonalizing the un-simplified $M^q$.
 \\

In \cite{Lee1986a, Lee1986b, Derman1979},
an $S_3$ symmetry was introduced among three fermion generations in the standard model,
resulting in an oversimplified $M^q$ pattern that has only two parameters in a quark type.
This simplification was achieved through the $S_3$ invariance of the Lagrangian.
However, in this model, two of the mass eigenvalues are degenerate, and the CKM matrix obtained is CP-conserving since both $U^q$ matrices are real.
This conflicts with the first necessary condition for achieving a complex $V_{CKM}$,
 which states that at least one of the $U^q$ matrices must be complex \cite{Lin2020a}. \\

In \cite{Lin1988, Lee1990}, the $S_3$-symmetric SM was extended with an extra Higgs doublet
which introduced a third parameter into $M^q$ and solved the mass-degeneracy problem.
However, such an $S_3$-symmetric 2HDM did not solve the CPV problem since the $U^u$ and $U^d$ matrices thus derived are the same,
$i.e.$, $U^u = U^d$, despite having some complex elements.
 Although they satisfy the first condition for a complex $V_{CKM}$ as stated in \cite{Lin2020a},
 they conflict with the second condition, which requires $U^u \neq U^d$.
These researches demonstrate that oversimplified $M^q$ patterns are inadequate for obtaining a satisfactory $V_{CKM}$.
To achieve a satisfactory solution to the CPV problem, we must search for a model with fewer and/or weaker constraints. \\

The researches mentioned above clearly indicate that the $S_3$ symmetry is too strong a constraint and oversimplifies the $M^q$ pattern, leading to a CP-conserving $V_{CKM}$.
This suggests that a more complex $M^q$ pattern with fewer and/or weaker constraints could provide a better solution to the CPV problem.
In \cite{Lin2020a}, three residual $S_2$ symmetries between two of the three fermion generations were considered,
resulting in three additional $M^q$ and $U^q$ patterns.
The extra $U^q$ patterns together with the $S_3$-symmetric one, provide us an opportunity to satisfy both necessary conditions stated in \cite{Lin2020a}.
Under certain conditions, CP symmetry was explicitly violated, with a Jarlskog invariant of $J\approx 0.171$, which is approximately 5,700 times larger than the one predicted by the current SM.
 \\

The original objective of \cite{Lin2020a} was to find matrix pairs that can be diagonalized simultaneously by the same $U^q$ matrix to address the FCNC problem in the 2HDM.
During the derivation, the natural-flavor-conserving (NFC) condition proposed by G. C. Branco \cite{Branco1985}, $M_1 \cdot M_2^{\dagger} -M_2 \cdot M_1^{\dagger}=0$, was frequently used.
However, as we obtained such matrix pairs, we could rearrange them as a product of a complex Yukawa-coupling matrix and the VEV of the unique Higgs doublet in the SM.
Subsequently, we obtain a standard model $M^q$ that violates CP symmetry "explicitly".
If the SM is already sufficient to solve these problems, we prefer to study them in the SM rather than in the 2HDM or other new physics. \\

 In \cite{Lin2019}, the $S_2$ symmetries were further removed away and only two assumptions were kept:
the Hermiticity of $M^q$ and a common $U^q$ matrix that diagonalizes the real and imaginary parts of $M^q$ simultaneously.
In \cite{Lin2021}, one of those two assumptions was further removed.
The researchers studied a naturally Hermitian matrix ${\bf M^2} = M^q_L \cdot M^{q\dagger}_L$,
rather than assuming that $M^q_L = M^{q\dagger}_L$ was Hermitian.
By definition, ${\bf M^2}=M^q_L \cdot M^{q\dagger}_L$ has the same $U^q_L$ matrix as $M^q_L$,
where the subindex $L$ denotes left-handed quarks, which is usually omitted if unnecessary.
The only assumption that remained was that the real and imaginary parts of ${\bf M^2}$ can be diagonalized by the same $U^q$ simultaneously.
Although this solution to the CPV problem is not the ultimate one, it is already very close to it, since the predicted CKM elements fit their corresponding empirical values to an accuracy of ${\bf O}(\lambda^{1/2})$ or better at tree level.
 \\

In \cite{Lin2021}, a very general pattern for ${\bf M^2}$ was given in terms of five parameters as follows:
\begin{eqnarray}
{\bf M^2} &=&
 \left( \begin{array}{ccc} {\bf A + B (x y- {x \over y})}  & {\bf y B}    & {\bf x B}   \\  {\bf y B}  & {\bf A +B ({y \over x}-{x \over y})}  &  {\bf B} \\ {\bf x B}  & {\bf B}                & {\bf A}   \end{array}\right)
+ i~\left( \begin{array}{ccc} 0     &  {\bf C \over y}     & - {\bf C \over x}   \\   - {\bf C \over y}      & 0  &  {\bf C} \\i {\bf C \over x}   & - {\bf C}  & 0   \end{array}\right) \nonumber \\
 &\equiv & {\bf M^2_R}+{\bf M^2_I},
\end{eqnarray}
where $\bf A$, $\bf B$, $\bf C$, $\bf x$, and $\bf y$ are composed solely of the Yukawa-couplings and the VEV of the SM Higgs doublet, as shown in Equations (6)-(14) of \cite{Lin2021}.
The eigenvalues of such a matrix were given analytically by
\begin{eqnarray}
{\bf m^2_1} &=& {\bf A-B {x \over y} -C {\sqrt{\bf x^2 +y^2 +x^2 y^2} \over {x y}}}, \nonumber \\
{\bf m^2_2} &=& {\bf A-B{x \over y} + C{\sqrt{\bf x^2 +y^2 +x^2 y^2} \over {x y}}},  \nonumber \\
{\bf m^2_3} &=& {\bf A+B{{(x^2+1) y} \over x}}.
\end{eqnarray}
\\

The unitary matrix ${\bf U^u}$ that simultaneously diagonalizes $\bf M^2_{Ru}$ and $\bf M^2_{Iu}$ for up-type quarks is given by:
\begin{eqnarray}
{\bf U^u} = \left( \begin{array}{ccc}
{-\sqrt{\bf x^2+y^2} \over \sqrt{\bf 2(x^2+y^2+x^2 y^2)}} & {\bf {x(y^2-i \sqrt{\bf x^2+y^2+x^2 y^2})} \over {\bf \sqrt{2} \sqrt{\bf x^2+y^2} \sqrt{\bf x^2+y^2+x^2 y^2}}} & {\bf {y(x^2+i \sqrt{\bf x^2+y^2+x^2 y^2})} \over {\bf \sqrt{\bf 2} \sqrt{\bf x^2+y^2} \sqrt{\bf x^2+y^2+x^2 y^2}}}  \\
 {-\sqrt{\bf x^2+y^2} \over \sqrt{\bf 2(x^2+y^2+x^2 y^2)}} & {{\bf x(y^2+i \sqrt{\bf x^2+y^2+x^2 y^2})} \over {\sqrt{\bf 2} \sqrt{\bf x^2+y^2} \sqrt{\bf x^2+y^2+x^2 y^2}}} &~{{\bf y(x^2-i \sqrt{\bf x^2+y^2+x^2 y^2})} \over {\sqrt{\bf 2} \sqrt{\bf x^2+y^2} \sqrt{\bf x^2+y^2+x^2 y^2}}} \\
 {{\bf x y} \over \sqrt{\bf x^2+y^2+x^2 y^2}} &~ {\bf y \over \sqrt{\bf x^2+y^2+x^2 y^2}} &~{\bf x \over \sqrt{\bf x^2+y^2+x^2 y^2}} \end{array}\right),  \nonumber \\
\end{eqnarray}
where all elements depend on only two of the five parameters.
Similarly, ${\bf U^d}$ for down-type quarks has the same pattern, and the parameters in it are denoted by primed symbols,
$\bf A'$, $\bf B'$, $\bf C'$, $\bf x'$, and $\bf y'$, respectively.
As a result,  a CKM matrix depending on four parameters, two from up-type quarks and two from down-type quarks, can be obtained.
With all the materials mentioned above collected, it becomes feasible to study how BAU was generated in such a standard model.
\\

\section{Variation of $R_{\Delta}~ \equiv ~{\Delta_{CP} ~/~ \Delta^{(0)}_{CP}}$ in the Natural parameters}

As mentioned in Section I, Jarlskog proposed a measure of CPV in the form of Equation (1).
However, the current value of this quantity is about ten orders of magnitude smaller than what is needed to account for the cosmologically observed BAU.
If we examine the components of $\Delta_{CP}$, it is clear that it is composed of a CKM-related Jarlskog invariant $J$ and two mass-related factors, which are defined as follows:
\begin{eqnarray}
\Delta m_{(u)}^2 &=& (m_t^2 - m_c^2 ) (m_c^2 - m_u^2 ) (m_u^2 - m_t^2 ) , \\
\Delta m_{(d)}^2 &=& (m_b^2 - m_s^2 ) (m_s^2 - m_d^2 ) (m_d^2 - m_b^2 ).
\end{eqnarray}
\\

By substituting the eigenvalues from Equation (5) into Equations (7) and (8), we obtain the expressions for the mass differences:
\begin{eqnarray}
\Delta m^2_{(u)} &=& {\bf 2 C [B^2 (x^2 +y^2 +x^2 y^2) - C^2] {{\bf (x^2 +y^2 +x^2 y^2)^{3/2}} \over {\bf x^3 y^3}}},  \\
\Delta m^2_{(d)} &=& 2 {\bf C'} [{\bf B'}^2 {\bf (x'^2 +y'^2 +x'^2 y'^2)}- {\bf C'}^2 ] {{\bf (x'^2 +y'^2 +x'^2 y'^2)^{3/2}} \over {\bf x'^3 y'^3}}.
\end{eqnarray}
Using these mass differences, we can rewrite Equation (1) as
\begin{eqnarray}
\Delta_{CP} &=&  J~  \Delta m^2_{(u)} ~\Delta m^2_{(d)} /T^{12}~.
\end{eqnarray}
However, the current value of $\Delta^{(0)}_{CP}$ is at most $\approx~ 10^{-20}$, far smaller than the value needed to explain the observed BAU,
as discussed in Section I and reported in \cite{Shaposhnikov1986}. \\

In \cite{Lin2020a}, it was found that in several $S_2$-symmetric cases,
 the value of $J$ is approximately 5,700 times larger than the value predicted by the SM.
However, even with such enhancements, the resulting value is still not enough to account for the $10^{10}$ discrepancy in the observed BAU.
It is worth noting that quark masses also contribute to $\Delta_{CP}$, and there are a total of ten free parameters in the equation.
However, due to the complexity of the equation, it is not feasible to study it analytically.
Therefore, we will consider a simplified example of the parameter sets to determine if it is possible to generate a large amount of BAU in this model.
\\

In \cite{Lin2021}, the authors pointed out the challenge of assigning which eigenvalue $m_i^2$ corresponds to which quark flavor $m_q^2$ (taking up-type quarks as an example), where $i=1,2,3$ denotes the eigenvalues and $q=u,c,t$ denotes the quark flavors.
There are thirty-six possible ways to make such assignments, but only eight of them were classified as being ${\bf O}(\lambda)$.
Moreover, only four of the eight ways were found to fit the empirical CKM elements to an accuracy of ${\bf O}(\lambda^{1/2})$ or better at tree level. \\

Then we denote those four $V_{CKM}$ candidates given in Equations (32) and (33) of \cite{Lin2021} by
\begin{eqnarray}
V[52] &=& V{\left( \begin{array}{ccc}  1 \\ 3 \\ 2 \end{array}\right) (2~3~1)} = \left( \begin{array}{ccc}   s    & p^*   & r^* \\  p'^*    & q    & p' \\ r & p    & s^* \end{array}\right) \nonumber \\
     = V^*[25] &=& V^*{\left( \begin{array}{ccc}  2 \\ 3 \\ 1 \end{array}\right) (1~3~2)}  , \\
 V[22] &=&  V{\left( \begin{array}{ccc}  2 \\ 3 \\ 1 \end{array}\right) (2~3~1)} = \left( \begin{array}{ccc} r  & p  & s^* \\  p'^*    & q    & p' \\ s & p^*    & r^* \end{array}\right)  \nonumber \\
     = V^*[55] &=& V^*{\left( \begin{array}{ccc}  1 \\ 3 \\ 2 \end{array}\right) (1~3~2)} ,
\end{eqnarray}
in which numbers in the square brackets denote their positions in $\bf Table~1$ of \cite{Lin2021} and the elements $p$, $p'$, $q$, $r$, and $s$ are also given in Equations (20)-(24) there.
It is trivial to find that $J$ of $V[25]$ and $V[22]$ are the same and those of $V[55]$ and $V[52]$ are the same.
Even more, $J$ of $V[22]$ and $V[55]$ have the same absolute value but opposite signs as what is shown below:
\begin{eqnarray}
J_{25}=J_{22} = {\rm Im}[p \cdot p \cdot r^* \cdot  s ] =-J_{55}=-J_{52},
\end{eqnarray}
if we take $V_{us}$, $V_{tb}$, $V^*_{ub}$ and $V^*_{ts}$ into consideration.
 \\

If we separate each of $p$, $r$, and $s$ into its real and imaginary parts as follows: $p=p_r +i p_i$, $r=r_r +i r_i$, and $s=s_r+ i s_i$,
it is straightforward to show that:
\begin{eqnarray}
{\rm Im} [ p \cdot p \cdot r^* \cdot s] &=& -{\rm Im}[p^* \cdot p^* \cdot r \cdot  s^*] \nonumber \\
 = (p^2_r -p^2_i) (r_r ~ s_i - r_i ~s_r ) &+& 2p_r ~ p_i  (r_r ~ s_r + r_i ~ s_i),
\end{eqnarray}
when taking $V_{us}$, $V_{tb}$, $V^*_{ub}$ and $V^*_{ts}$ into account.\\

Substituting Equations (9), (10), and (15) into (11) yields a complete expression for $\Delta_{CP}$ in terms of ten natural parameters.
Next, we can vary these parameters from their current values to some $S_N$-symmetric values and compare the resulting $\Delta_{CP}$ with its current value $\Delta^{(0)}_{CP}\approx 10^{-20}$ throughout the process.
\\

Subsequently, several tests were performed on the derived CKM elements by fitting them with empirical data,
and it was found that there were sixty-four sets of $\bf x$, $\bf y$, $\bf x'$, and $\bf y'$ that gave the same CKM element values.
If we substitute these values and the current quark masses into Equation (11),
we obtain correspondingly sixty-four sets of $\bf A$, $\bf B$, $\bf C$, $\bf A'$, $\bf B'$, and $\bf C'$ values.
The quark masses used here are
\begin{eqnarray}
m^{(0)}_u &=& 0.0023~{\rm GeV},~m_c^{(0)} =1.275~{\rm GeV},~m^{(0)}_t =173.2~{\rm GeV}, \nonumber \\
m^{(0)}_d &=& 0.0048~{\rm GeV},~m_s^{(0)} =0.095~{\rm GeV},~m^{(0)}_b =4.18~{\rm GeV},
\end{eqnarray}
where $m^{(0)}_q$ are current quark masses for $q=~u,~c,~t,~d,~s$, and $b$.
Using the current values of the CKM matrix elements and quark masses, we obtain a value of approximately $3.19 \times 10^9$ for the mass-squared part of Equation (11), denoted as $\Delta m^{(0)2}_{(u)} \cdot \Delta m^{(0)2}_{(d)}$,
We will compare this with the corresponding value obtained when the quark masses are varied according to the $S_N$-symmetric parameters. \\

In this way, we obtain sixty-four sets of $\bf x$, $\bf y$, $\bf x'$, $\bf y'$, ${\bf A,~B,~C,~A',~B'}$, and ${\bf C'}$ values.
However, half of them are excluded because some of the masses of down-type quarks are predicted to be imaginary at the $S_2$-symmetric point ${\bf (x',~y')}$= (-1, 1), which is obviously irrational.
As a result, only thirty-two of them are presented in $\bf TABLE~1$. \\

If all ten parameters vary arbitrarily,
it's theoretically always possible to find parameter spaces in which $\Delta_{CP}$ is tens orders stronger than $\Delta^{(0)}_{CP} \approx 10^{-20}$.
However, it is really too arbitrary to have any physical meanings.
Thus, in what follows we will make assumptions on some of the parameters and concentrate the discussions on one very simplified case. \\

Firstly, we assume that $\bf A$, $\bf B$, $\bf C$, $\bf A'$, $\bf B'$, and $\bf C'$ are fixed or very slowly varying quantities during the $S_2$-breaking process to be studied below, so we may use their current values listed in $\bf TABLE~1$ in the following calculations.
 Secondly, we assume that the residual $S_2$ symmetries for up- and down-type quarks were not broken down simultaneously,
  and that it happened among up-type quarks first while the down-type ones still possess the $S_2$ symmetry.
As mentioned in \cite{Lin2020a}, there are three $S_2$-symmetric patterns, $\bf x'=-y'=-1$, $\bf x'=-y'=1$, and $\bf x'=y'=-1$.
 Here, we select only the $\bf x'=-y'=-1$, or ($\bf x',~y'$)= (-1, 1), case for an example in the following calculations. \\

The proposed method for examining parameter spaces for an extremely large $\Delta_{CP}$ is to numerically test the values of $\bf x$ and $\bf y$ from their $S_2$-symmetric values ($\bf x,~ y$)= (-1, 1) to their current values.
To quantify the magnitude of $\Delta_{CP}$ in different parameter spaces,
the ratio of the running $\Delta_{CP}$ to its current value $\Delta^{(0)}_{CP}$ is defined as:
\begin{eqnarray}
R_{\Delta}= {\Delta_{CP} \over \Delta^{(0)}_{CP}}={{J\cdot \Delta m^2_{(u)} \cdot \Delta m^2_{(d)}}\over {J^{(0)}\cdot \Delta m^{(0)2}_{(u)} \cdot \Delta m^{(0)2}_{(d)}}},
\end{eqnarray}
in which a common factor $T^{12}$ is canceled out naturally.
In this way, putting $T^{12}$ or $v^{12}$ into Equation (1) will not affect not a bit the final result.
 \\

If we examine $R_{\Delta}$ by letting $\bf x$ and $\bf y$ to run from ($\bf x,~ y$)= (-1, 1) to their current values given in $\bf TABLE~1$,
the three-dimensional (3D) plots of all thirty-two candidate sets demonstrate many parameter spaces in which $R_{\Delta}~\ge~10^{10}$.
This means that under such circumstances, CPV could be extremely strong, leading to the generation of a large amount of BAU.
\\

As shown in FIG. 1, all thirty-two 3D plots exhibit many areas in which $R_{\Delta}~\ge~10^{10}$.
In some of them, we observe a peak emerging from the green $R_{\Delta}=10^{10}$ plane near the point (${\bf x,~y}$)= (-1, 1).
In fact, such a peak always exists in all thirty-two plots.
The reason why we do not see them in cases 01, 04, 14, and 15 is that they are obscured by the plotting ranges of $\bf x$ and $\bf y$ in the graphics software.
However, if we zoom in on the region around (${\bf x,~y}$)= (-1, 1), as shown in FIG. 2, the peak becomes prominent.
 \\

It is worth noting that in all thirty-two cases, $R_{\Delta}$ becomes extremely large,
on the order of $10^{10}$, as the values of ${\bf x}$ and ${\bf y}$ approach the $S_2$-symmetric point (${\bf x,~y}$)= (-1, 1).
However, at that exact point, $R_{\Delta}$ is equal to 0 since $p_i = r_i = s_i = 0$,
and therefore $J = 0$ if we substitute (${\bf x,~y}$)= (-1, 1) into Equation (17).
This discontinuity at the $S_2$-symmetric point indicates that $R_{\Delta}$ is highly sensitive to small variations in the values of ${\bf x}$ and ${\bf y}$ near that point. \\

To summarize, all thirty-two cases studied in this article have regions where $R_\Delta\gg10^{10}$,
 indicating the possibility of extremely strong CPV that could generate the observed baryon asymmetry in the universe.
These regions are often concentrated around the $S_2$-symmetric point (${\bf x,~y}$)= (-1, 1).
It is suggested that the evolution of the universe, from a state where both up- and down-type quarks were $S_2$-symmetric to a state where only the down-type quarks remained $S_2$-symmetric, may have caused a variation in the strength of CPV.
This variation in CPV strength could have been exceptionally strong in certain parameter spaces,
 resulting in the generation of the observed baryon asymmetry in the universe.
It also suggests that CP symmetry can be violated in conjunction with the breakdown of $S_N$ symmetries,
at least in the scenario presented in this article.
This provides a potential solution to both the CPV problem and the BAU problem without introducing additional complications.
 \\

\section{Conclusions and Discussions}

This article explores the production of the BAU in the SM by studying the Jarlskog measure of CP violation, $\Delta_{CP}$.
Previous research has shown that the SM alone is already enough to ignite CPV explicitly and the results fit experiments to an accuracy of ${\bf O}(\lambda^{1/2})$ or better, and that six quark masses are composed of ten natural parameters,
 while the nine CKM elements are composed of only four of these parameters.
These parameters are considered "natural" since they are solely composed of the Yukawa couplings and the vacuum expectation value $v$.
Such a parameterization of the CKM matrix in four parameters is as natural as the Kobayashi-Maskawa parameterization \cite{KM1973} or the Standard parameterization \cite{Chau1984}, which are similar to different coordinate systems in geometry.
This provides us with a new perspective to investigate the relationships between the CKM matrix
 and the Yukawa coupling matrices that is not present in other parameterizations. \\

By examining thirty-two sets of $\bf x$, $\bf y$, $\bf x'$, and $\bf y'$ that have the same CKM performances,
we were able to obtain thirty-two sets of the parameters $\bf A$, $\bf B$, $\bf C$, $\bf A'$, $\bf B'$, and $\bf C'$.
In order to investigate their potential for producing BAU,
we simplified the problem by assuming that six of the parameters are constants and that two are fixed by assuming an $S_2$ symmetry in the down-quark sector.
To measure the performance of $\Delta_{CP}$ in the remaining free parameters $\bf x$ and $\bf y$, we introduced a ratio $R_{\Delta} \equiv \Delta_{CP} /\Delta^{(0)}_{CP}$. \\

We then allowed $\bf x$ and $\bf y$ to vary from the $S_2$-symmetric point $({\bf x, y})=(-1,1)$ to their current values and found regions in all thirty-two cases where $R_{\Delta}$ is much larger than $10^{10}$,
providing evidence that SM alone can produce a significant amount of BAU.
Interestingly, we also observed that in all thirty-two plots, $R_{\Delta}$ diverges as the parameters approach the point $({\bf x, y})=(-1,1)$, where $J=0$ and $R_{\Delta}=0$.
This suggests a discontinuity at that point.  \\

It is important to note that while the investigation presented here is simplified,
it still provides valuable insights into the CPV and BAU problems within the SM.
Furthermore, the natural parametrization derived here is a significant contribution to the field and can potentially lead to further advancements in this area of research.
It is also important to continue exploring other candidate states and further refining the model in order to better understand the complexities of these problems.
Ultimately, continued research in this area will help us to gain a deeper understanding of the universe and the fundamental processes that govern it. \\

\begin{figure}
  \begin{center}
   \includegraphics{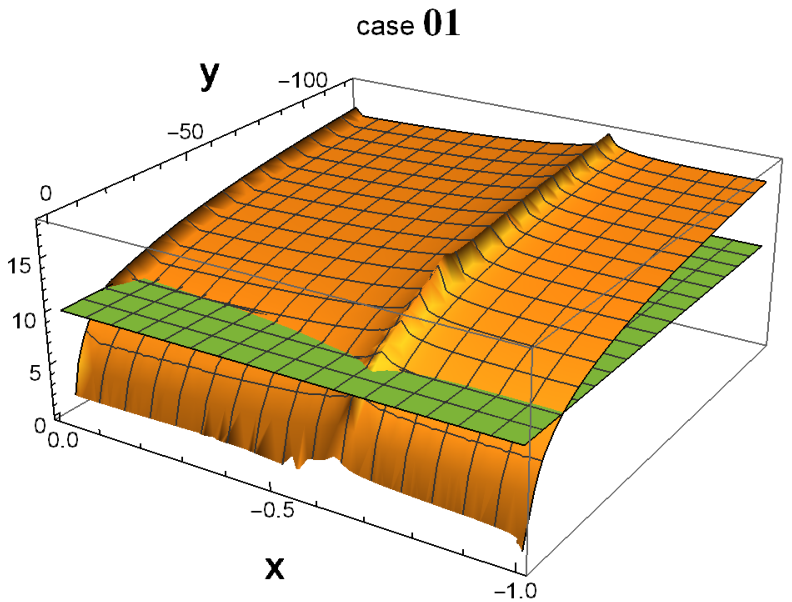}

   \includegraphics{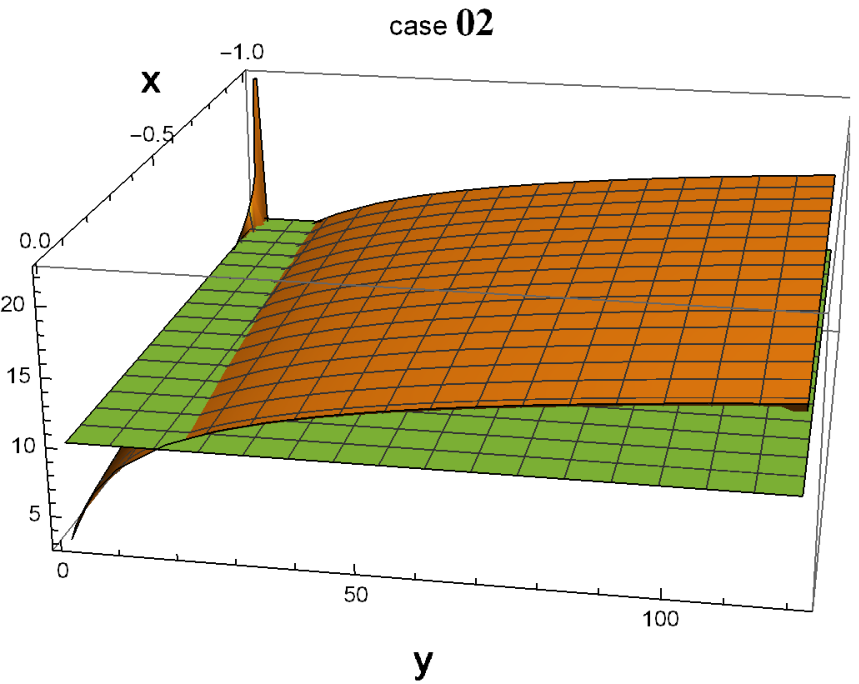}

   \includegraphics{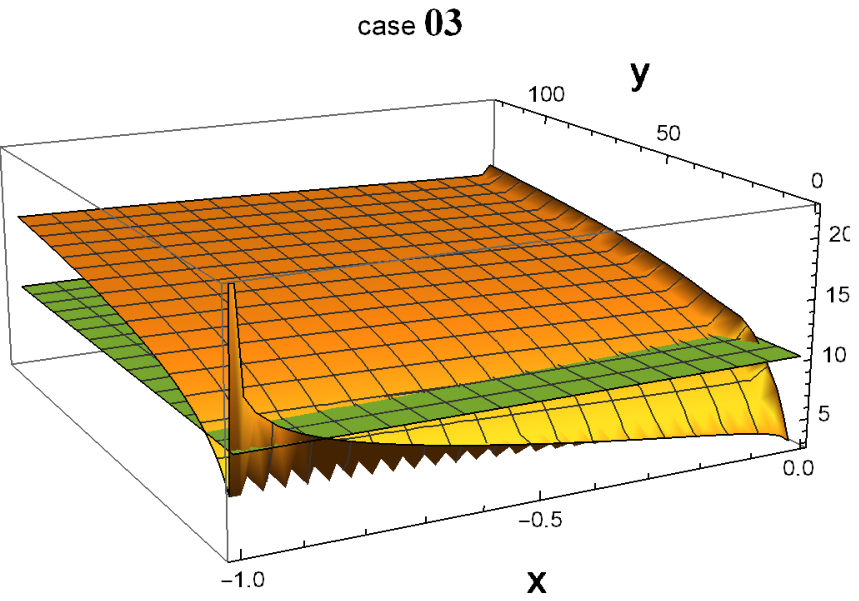}

   \includegraphics{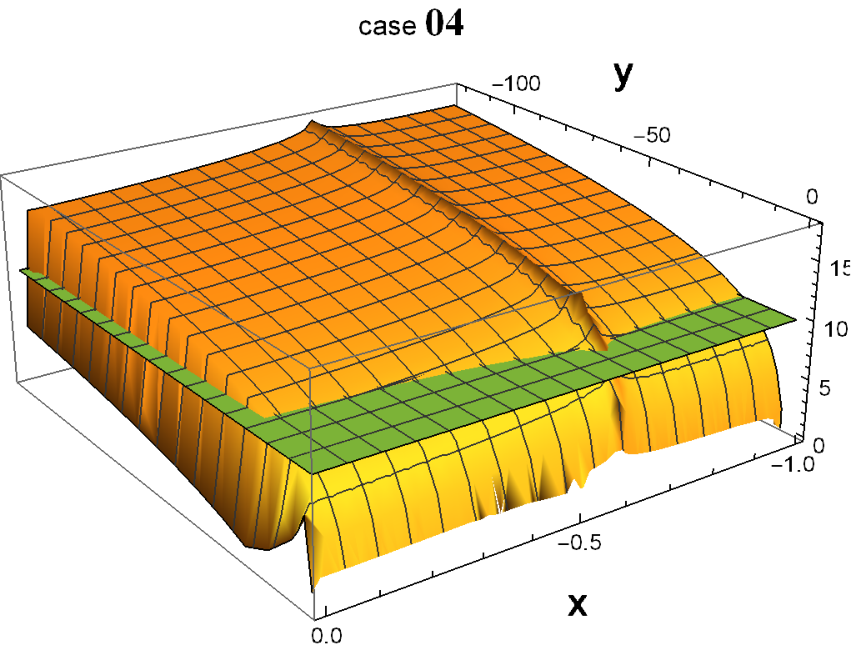}
   \end{center}
\end{figure}

\begin{figure}
  \begin{center}
   \includegraphics{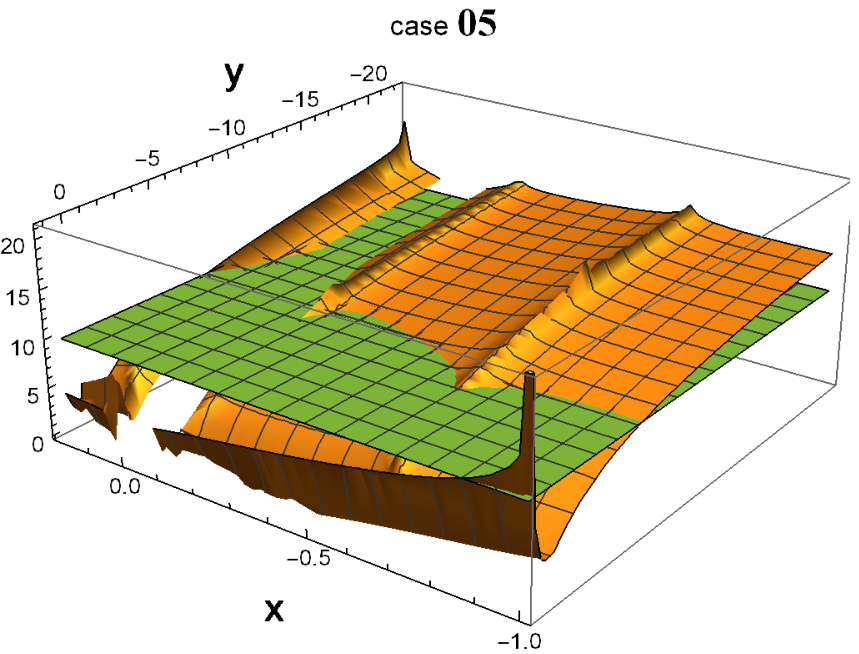}

   \includegraphics{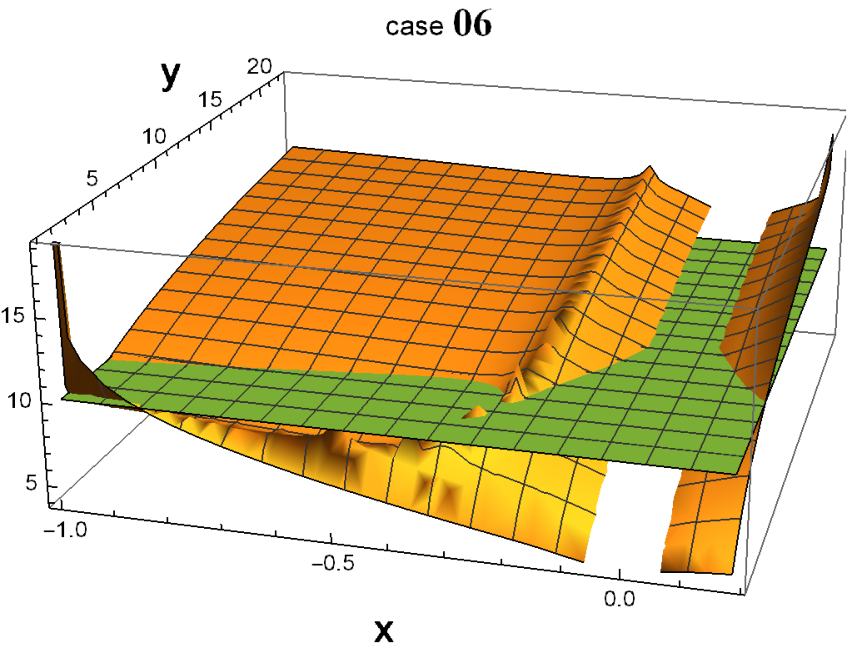}

   \includegraphics{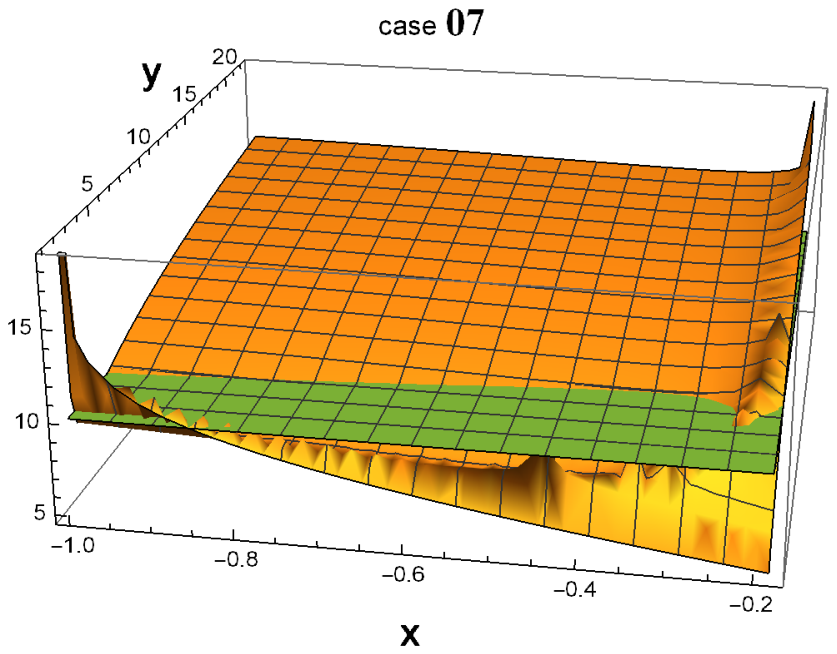}

   \includegraphics{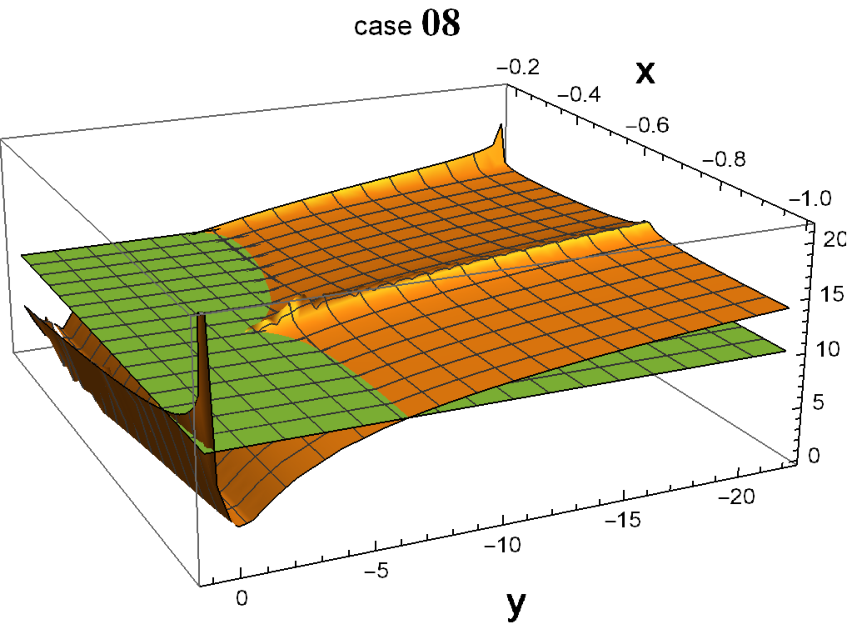}
   \end{center}
\end{figure}

\begin{figure}
  \begin{center}
   \includegraphics{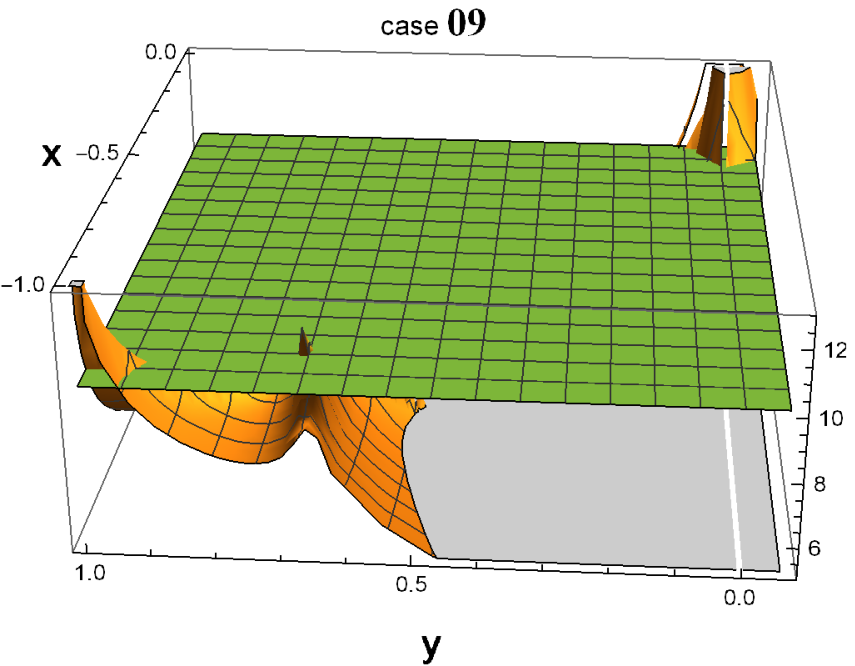}

   \includegraphics{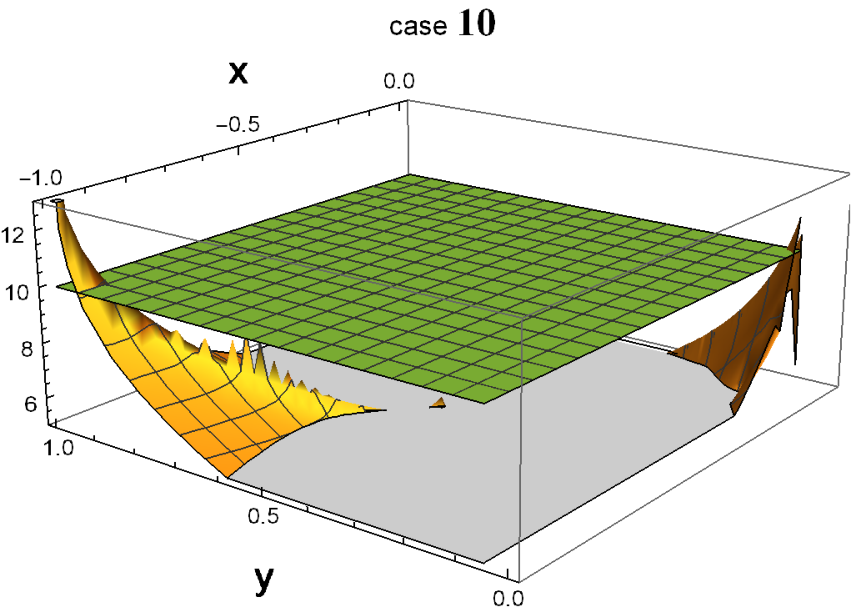}

   \includegraphics{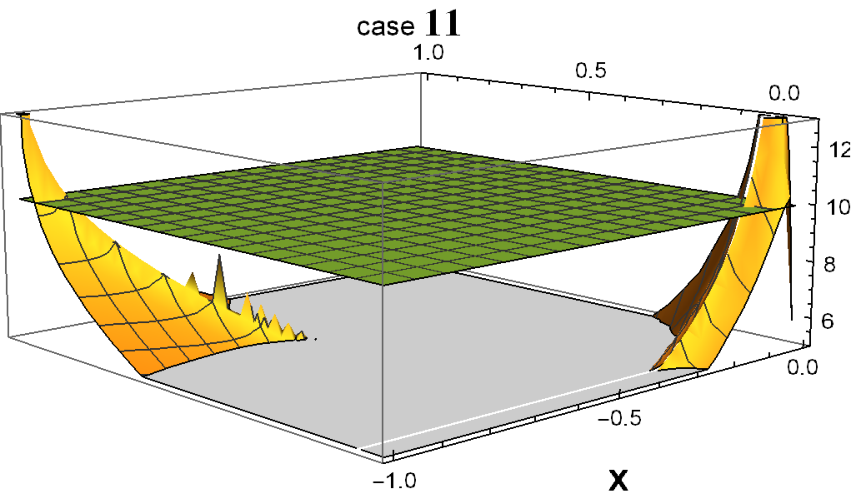}

   \includegraphics{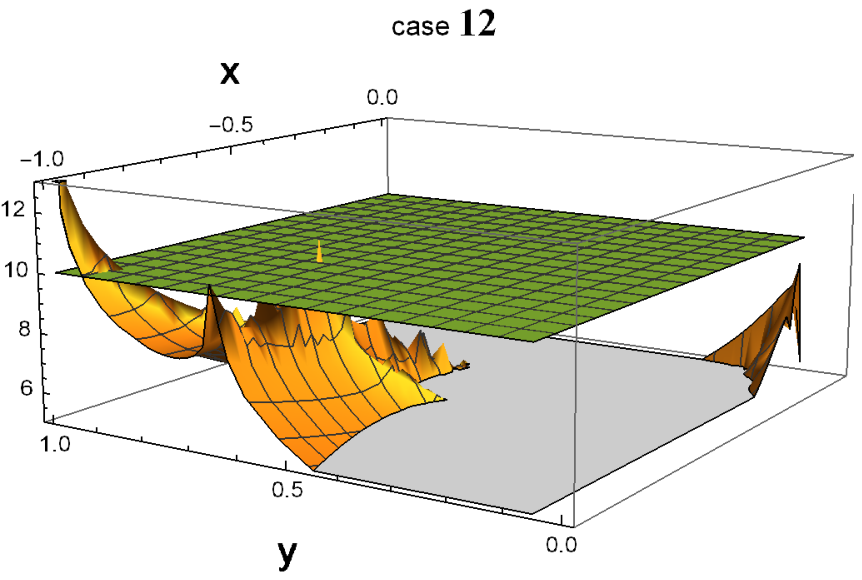}
   \end{center}
\end{figure}

\begin{figure}
  \begin{center}
   \includegraphics{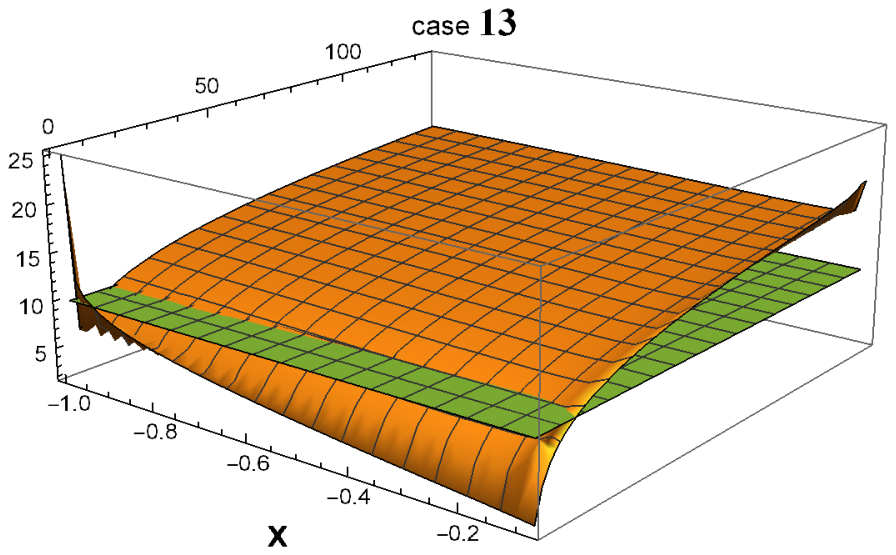}

   \includegraphics{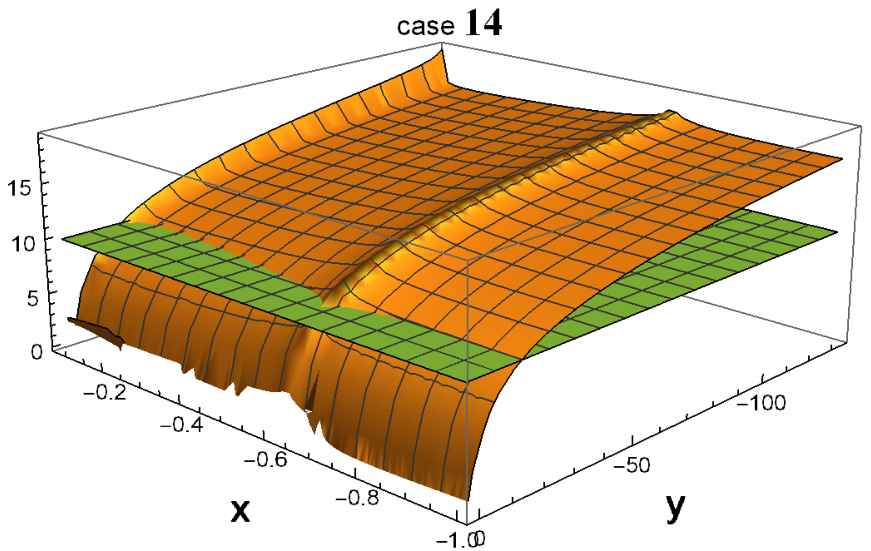}

   \includegraphics{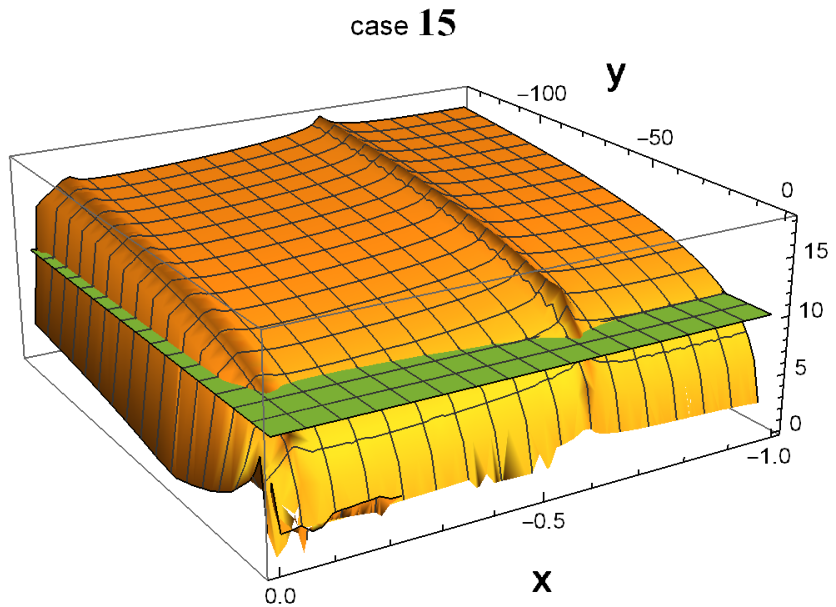}

   \includegraphics{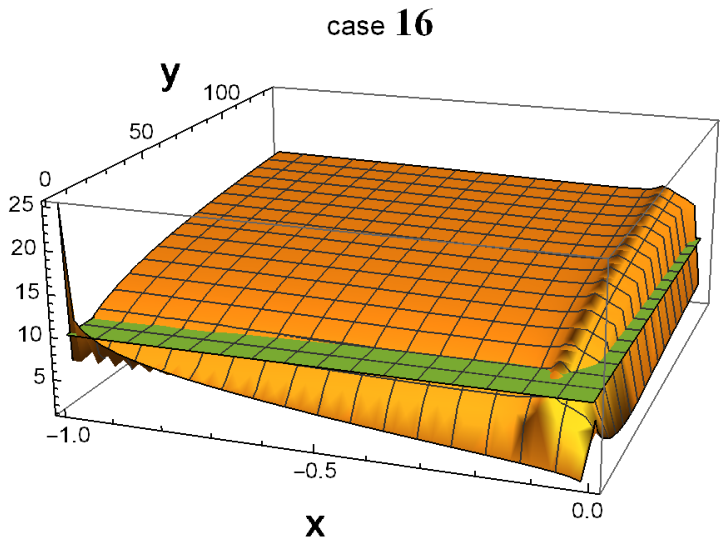}
      \end{center}
\end{figure}

\begin{figure}
  \begin{center}
   \includegraphics{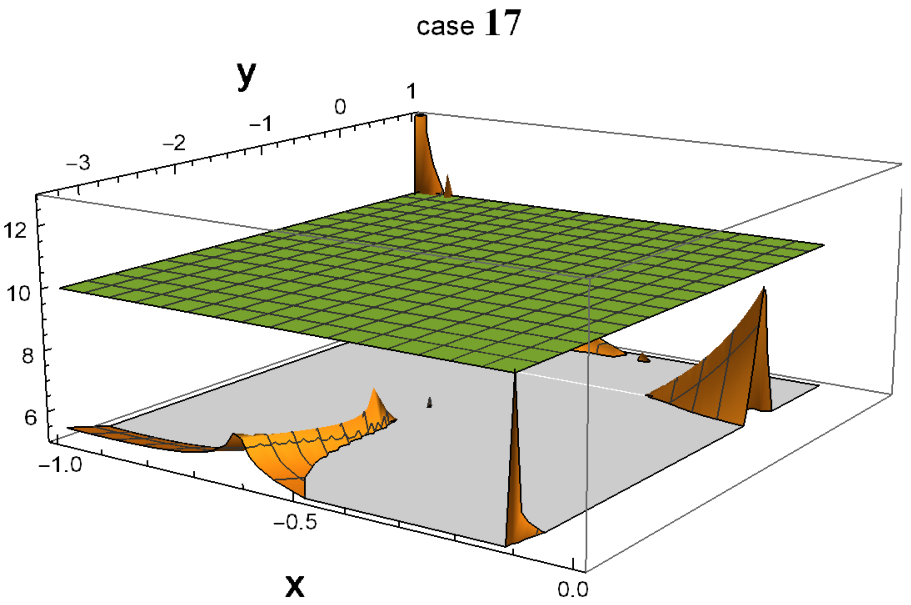}

   \includegraphics{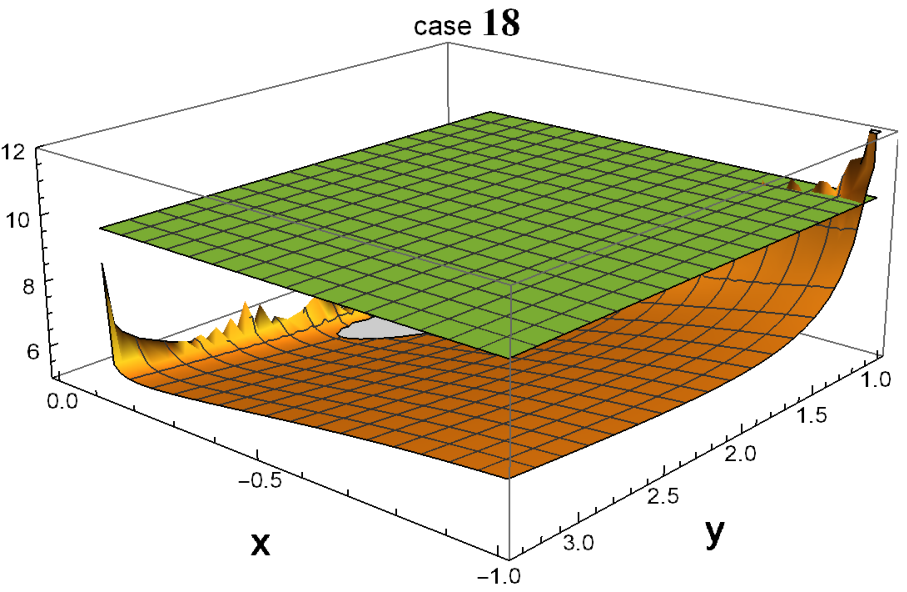}

   \includegraphics{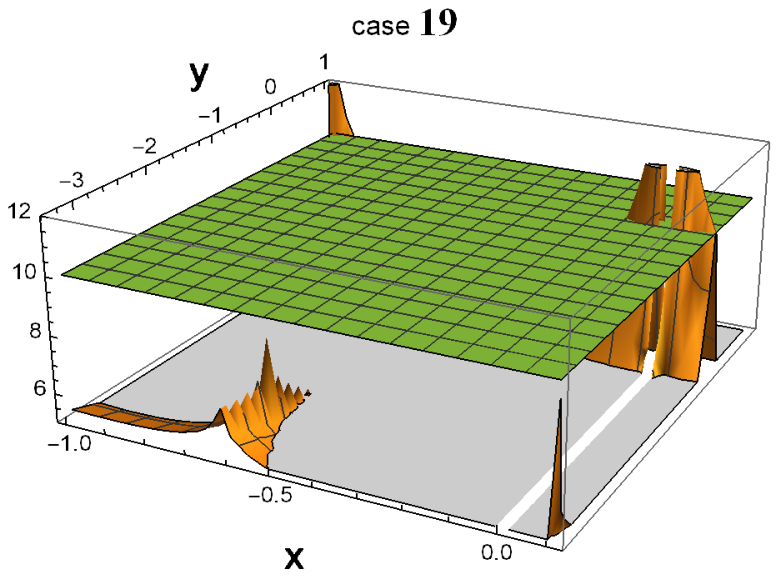}

   \includegraphics{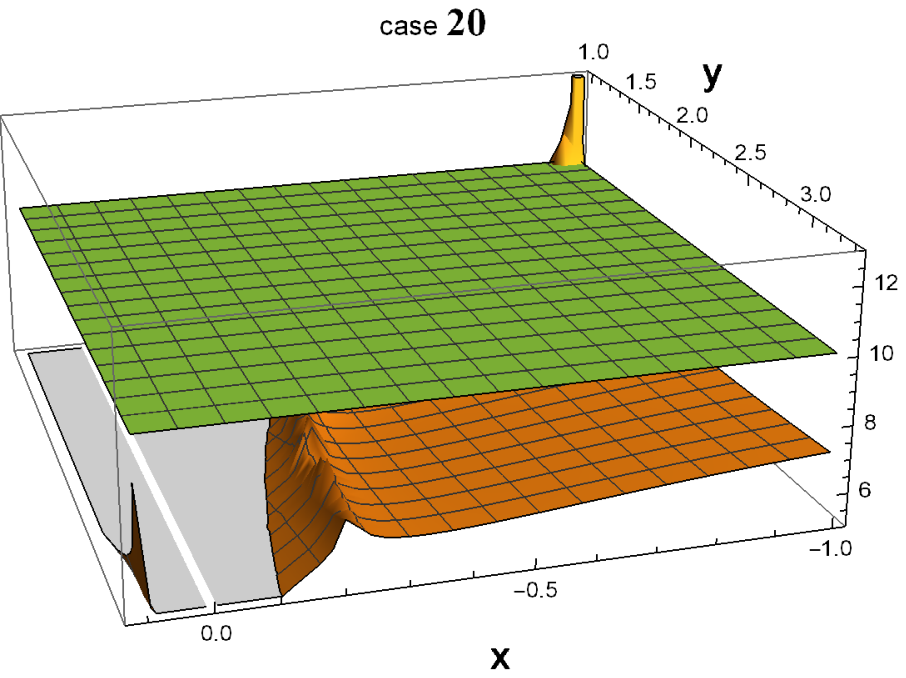}
   \end{center}
\end{figure}

\begin{figure}
  \begin{center}
   \includegraphics{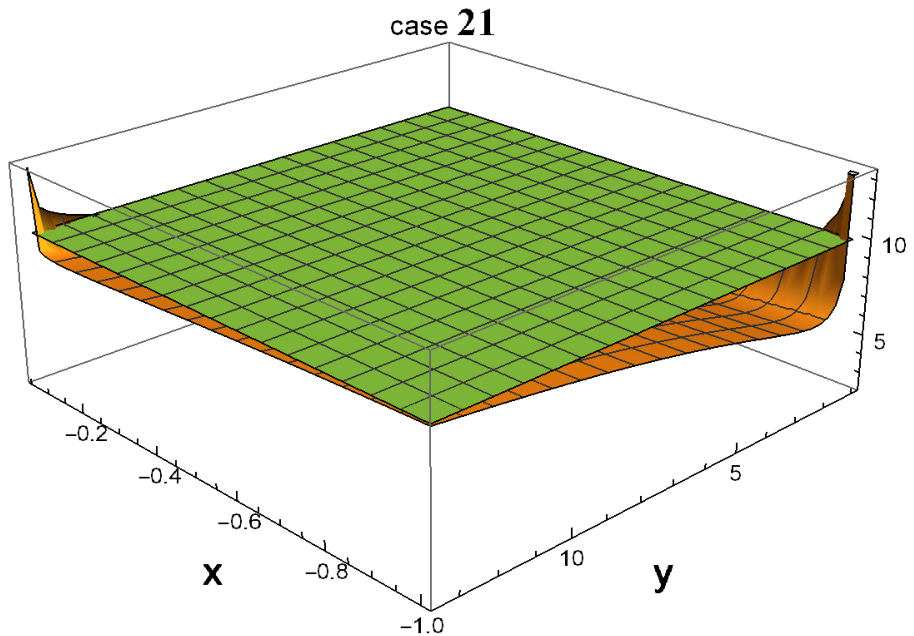}

   \includegraphics{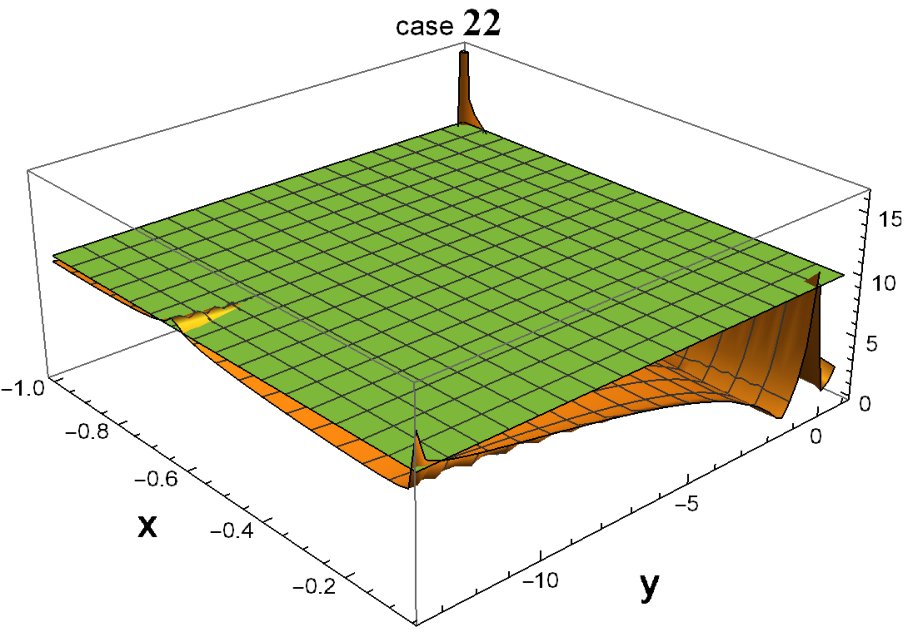}

   \includegraphics{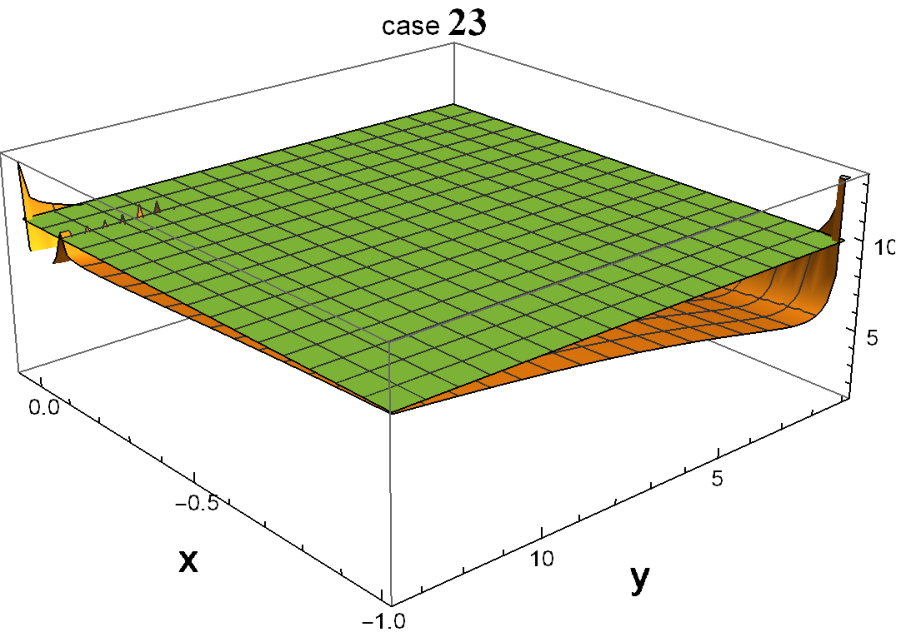}

   \includegraphics{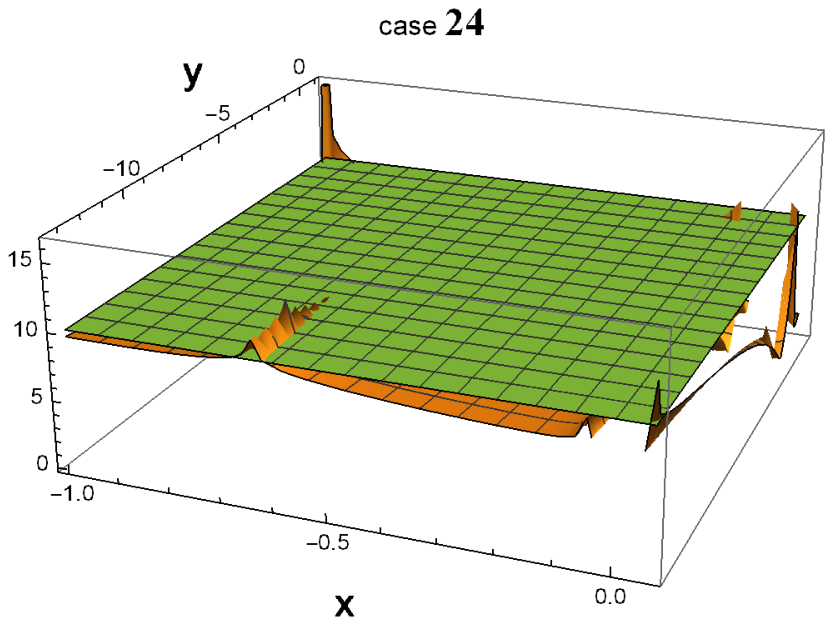}
      \end{center}
\end{figure}

\begin{figure}
  \begin{center}
   \includegraphics{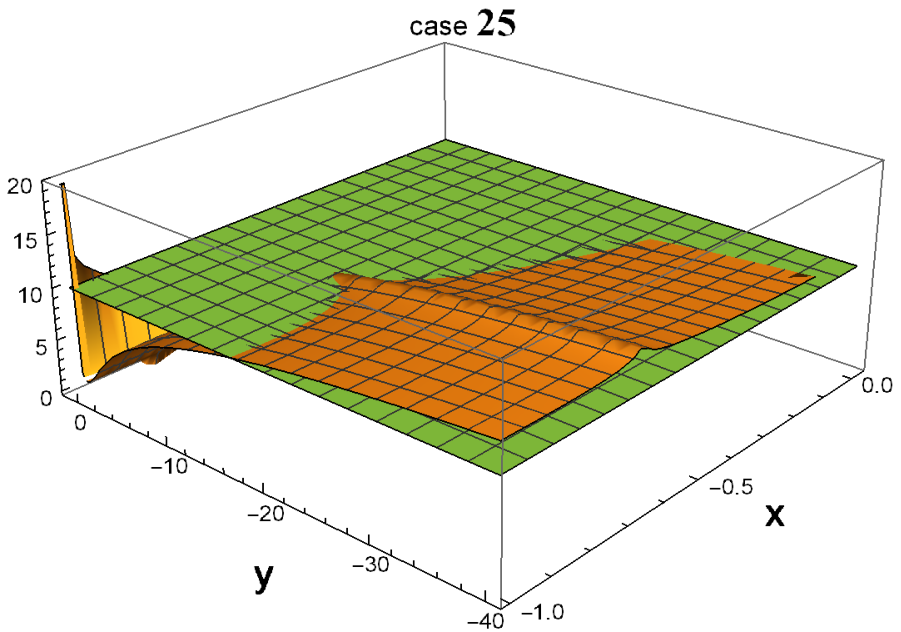}

   \includegraphics{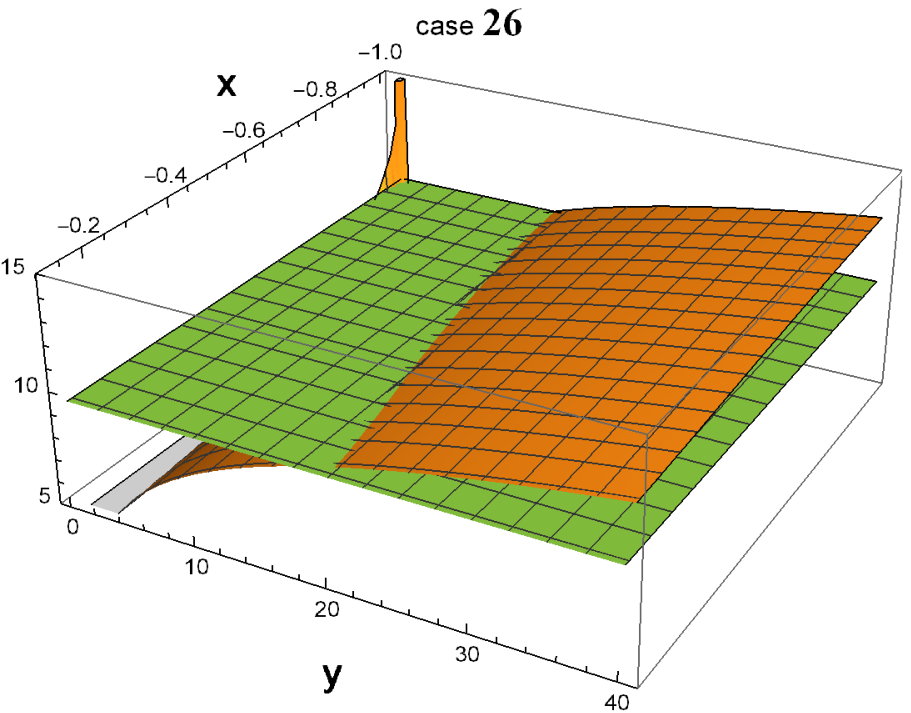}

   \includegraphics{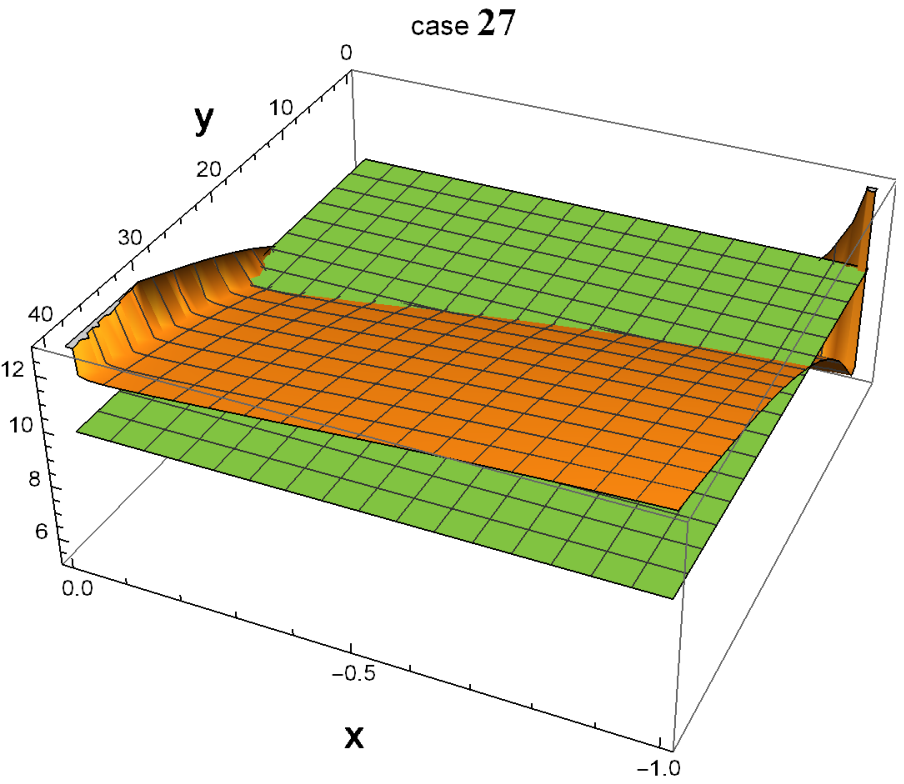}
   \end{center}
\end{figure}

\begin{figure}
  \begin{center}
   \includegraphics{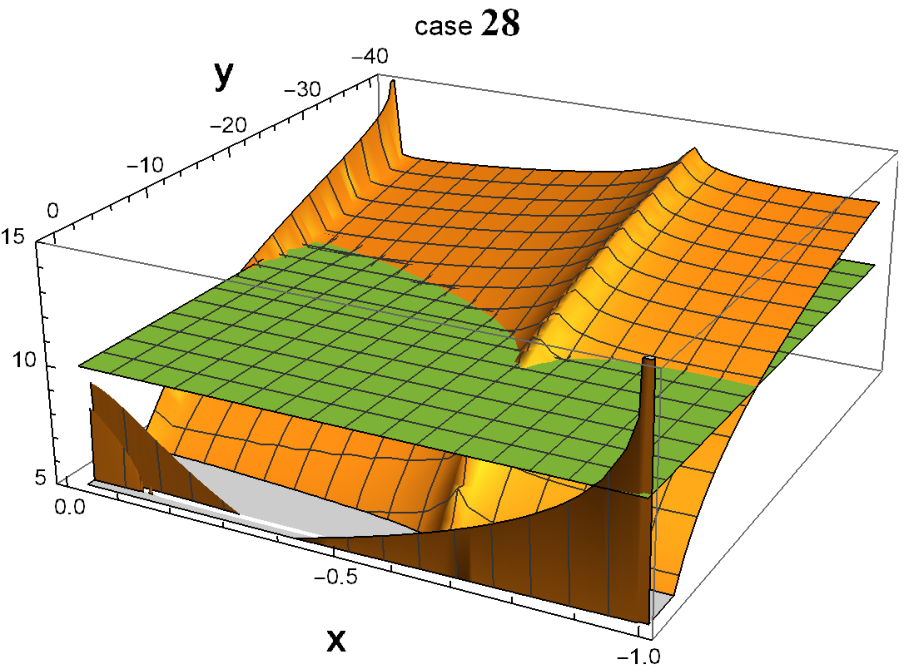}

   \includegraphics{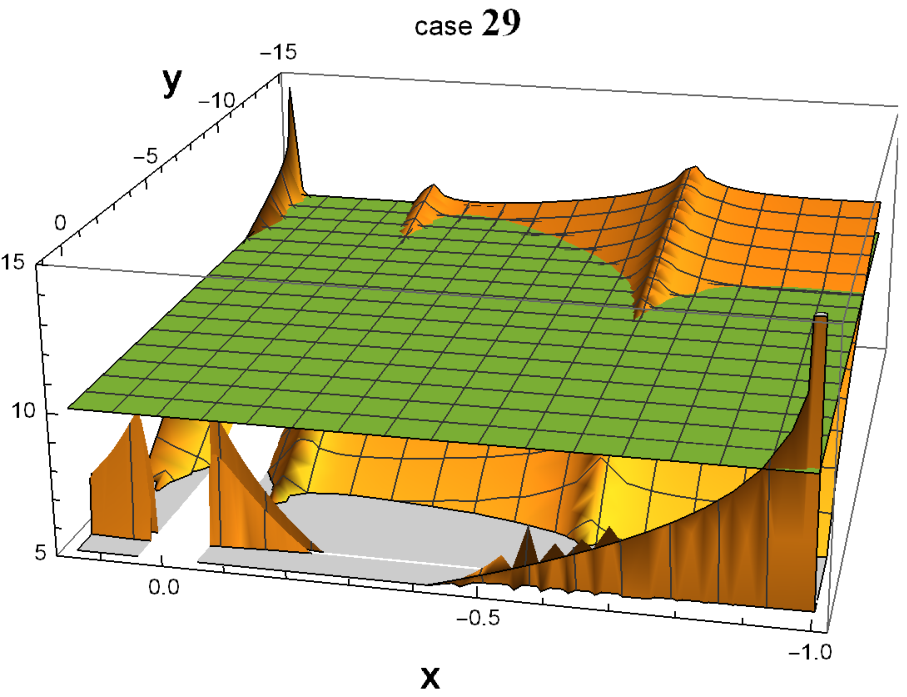}

   \includegraphics{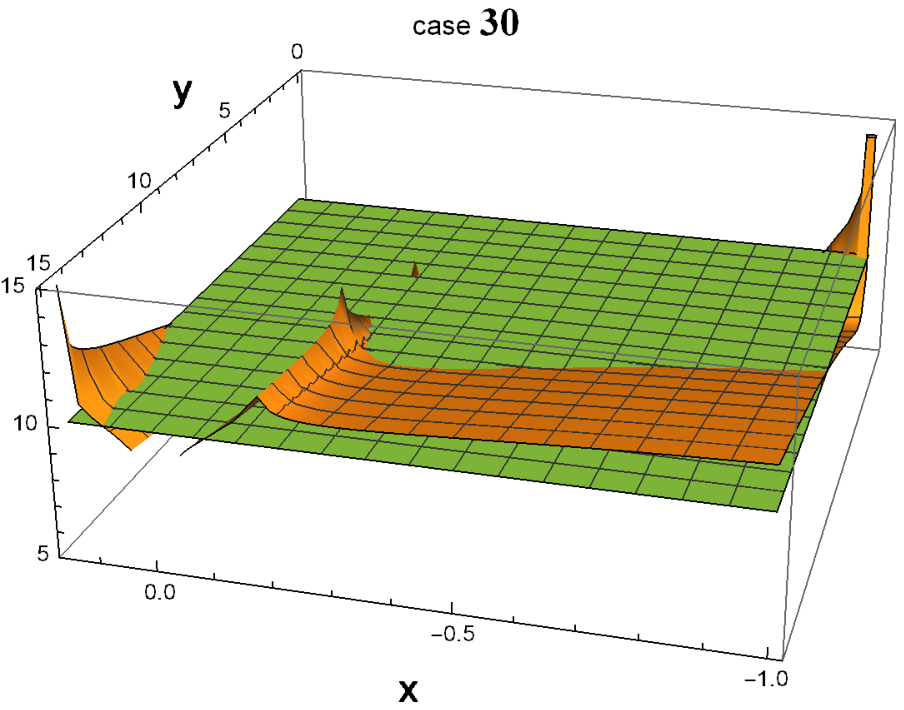}
      \end{center}
\end{figure}

\begin{figure}
  \begin{center}
   \includegraphics{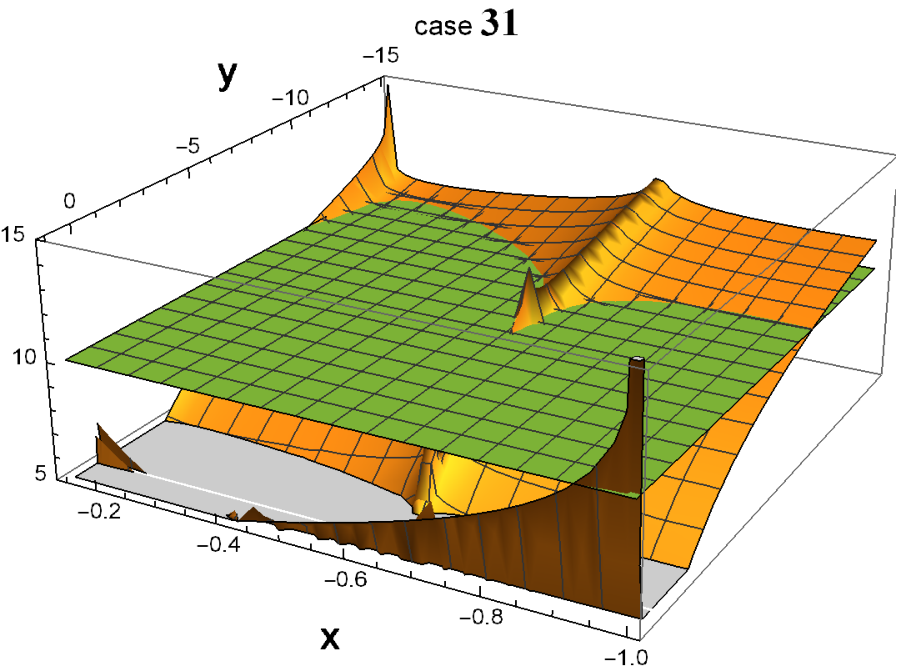}

   \includegraphics{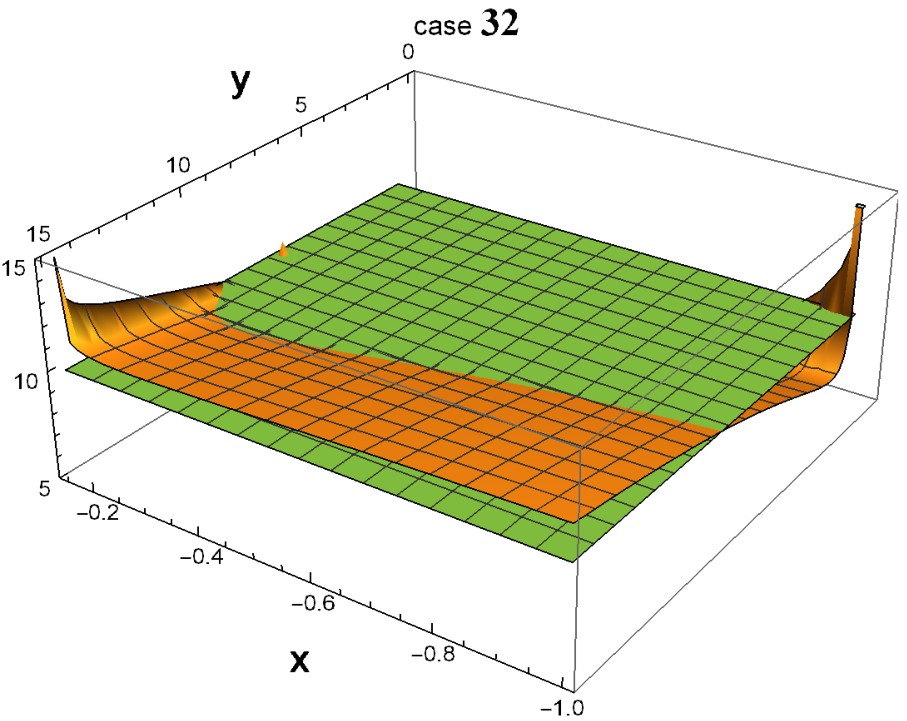}
      \end{center}
      \caption{\label{tab:i}
     The 3D plots depict the values of $R_{\Delta}$ in the $\bf x$ and $\bf y$ parameter spaces, with the z-axis representing the logarithm (base 10) of $R_{\Delta}$.
     The parameter $\bf x$ ranges from -1 to its current value, as given in $\bf TABLE~1$, while $\bf y$ ranges from 1 to its current value.
     Notably, the plots reveal several parameter spaces where $R_{\Delta}$ is significantly greater than $10^{10}$, particularly around the point ($\bf x, ~y$)=(-1, 1).
     This suggests that regardless of which scenario actually occurred in our universe,
     there must have been a highly productive era when the $S_2$ symmetry of down-type quarks was broken.  }
\end{figure}

\begin{figure}
  \begin{center}
   \includegraphics{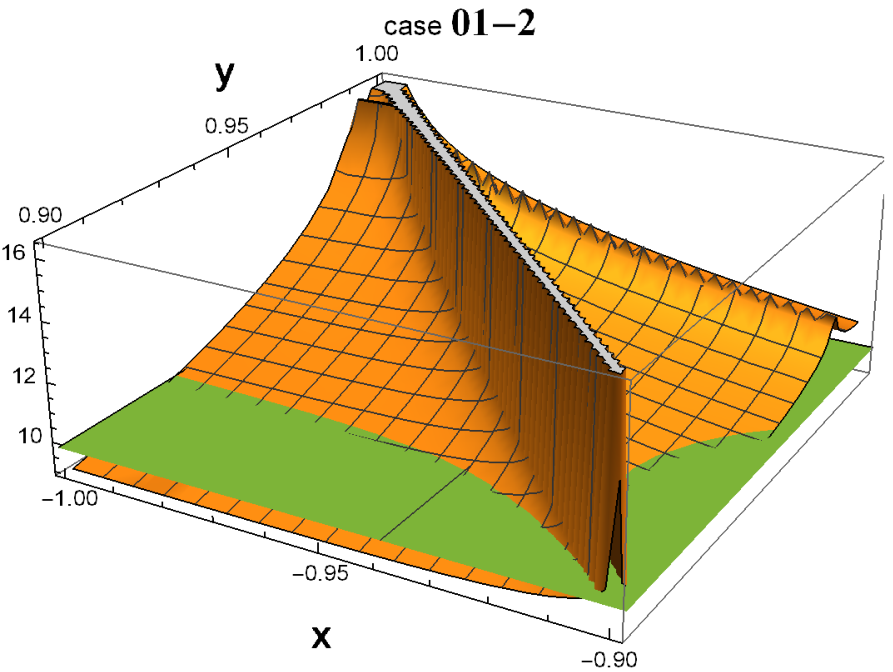}

   \includegraphics{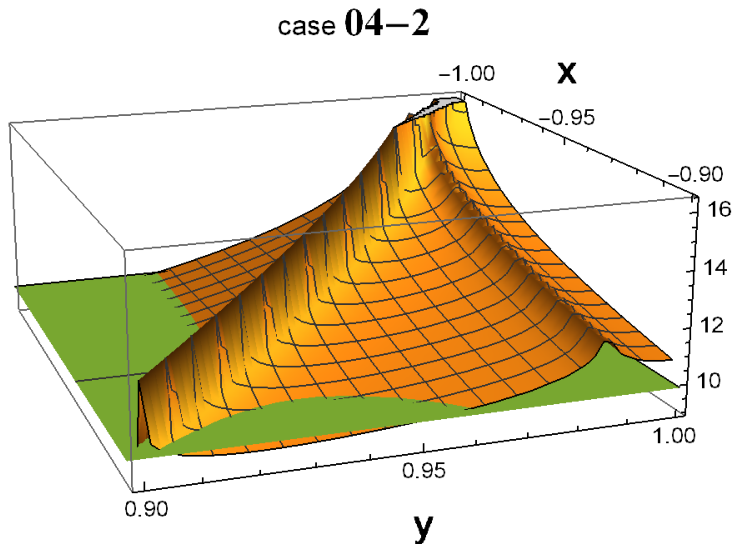}

   \includegraphics{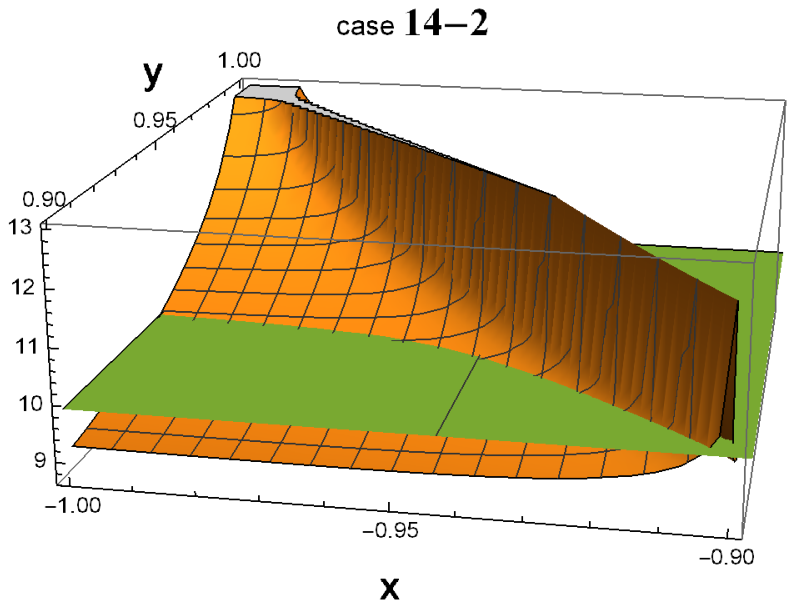}

   \includegraphics{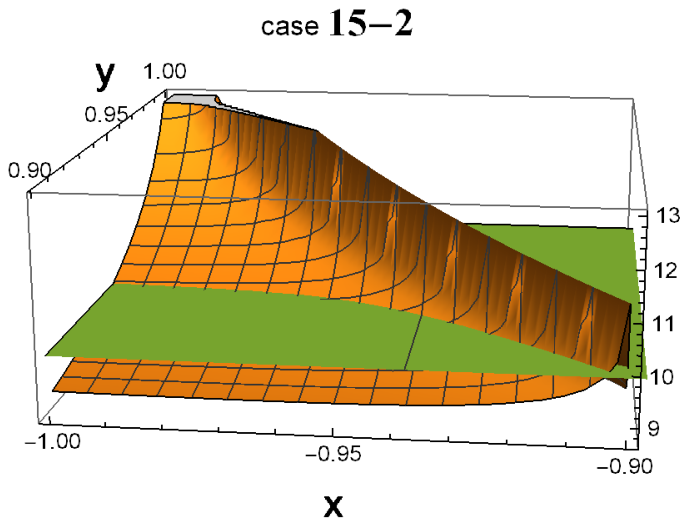}
         \end{center}
         \caption{\label{tab:i}
         For cases 01, 04, 14, and 15, the peaks that extend beyond the $R_{\Delta} = 10^{10}$ plane around the point ($\bf x, ~y$)=(-1, 1) are not particularly conspicuous.
          To obtain a clearer view, we zoom in on this area in the four corresponding figures. }
\end{figure}

\begin{table*}[tbp]
\begin{tabular}{|l|llllllllll|}
\hline
$\#$ & ~~~~$\bf x$ & ~~~~$\bf y$ & ~~~~$\bf x'$ & ~~~~$\bf y'$ & ~~~~$\bf A$ & ~~~~$\bf B$ & ~~~~$\bf C$ & ~~~~ $\bf A'$ & ~~~~$\bf B'$ & ~~~~$\bf C'$  \\
\hline
$~$1 & -0.00744658	& -121.51 & -0.182684 & $~$22.198 & $~$14923 & -0.914387 & $~$111.122 & $~$8.73564 & $~$0.0695014 & -1.56993  \\
$~$2 & -0.00744658	& $~$121.51 & -0.182684 & -22.198 & $~$14923 & $~$0.914387 & -111.122 & $~$8.73564 & -0.0695014 & $~$1.56993  \\
$~$3 & $~$0.00744658 & $~$121.51 & $~$0.182684 & -22.198 & $~$14923 & -0.914387 & $~$111.122 & $~$8.73564 & $~$0.0695014 & -1.56993  \\
$~$4 & $~$0.00744658 & -121.51 & $~$0.182684 & $~$22.198 & $~$14923 & $~$0.914387 & -111.122 & $~$8.73564 & -0.0695014 & $~$1.56993  \\
\hline
$~$5 & $~$0.182684 & -22.198   & $~$0.00744658 & $~$121.51 & $~$14922   & $~$118.826  & -2681.72 & $~$8.73621 & -0.000534827 & $~$0.0650529  \\
$~$6 & $~$0.182684  & $~$22.198 & $~$0.00744658 & -121.51   & $~$14922   & -118.826 & $~$2681.72 & $~$8.73621 & $~$0.000534827 & -0.0650529  \\
$~$7 & -0.182684  & $~$22.198 & -0.00744658   & -121.51   & $~$14922   & $~$118.826 & -2681.72 & $~$8.73621 & -0.000534827 & $~$0.0650529  \\
$~$8 & -0.182684  & -22.198   & -0.00744658   & $~$121.51 & $~$14922   & -118.826 & $~$2681.72 & $~$8.73621 & $~$0.000534827 & -0.0650529  \\
\hline
$~$9 & $~$0.00909528 & -0.0608101 & -0.082085 & $~$134.29 & $~$14596.5 & $~$2182.76 & -134.23 & $~$8.73621 & $~$0.00529903 & -0.714706  \\
10 & $~$0.00909528 & $~$0.0608101 & -0.082085 & -134.29 & $~$14596.5 & -2182.76 & $~$134.23 & $~$8.73621 & -0.00529903 & $~$0.714706  \\
11 & -0.00909528 & -0.0608101 & $~$0.082085 & $~$134.29 & $~$14596.5 & -2182.76 & $~$134.23 & $~$8.73621 & -0.00529903 & $~$0.714706  \\
12 & -0.00909528 & $~$0.0608101 & $~$0.082085 & -134.29 & $~$14596.5 & $~$2182.76 & -134.23 & $~$8.73621 & $~$0.00529903 & -0.714706  \\
\hline
13 & -0.082085 & $~$134.29 & $~$0.00909528 & -0.0608101 & $~$14923 & $~$9.05968 & -1220.85 & $~$8.54526 & $~$1.2767 & -0.0785808  \\
14 & -0.082085 & -134.29 & $~$0.00909528 & $~$0.0608101 & $~$14923 & -9.05968 & $~$1220.85 & $~$8.54526 & -1.2767 & $~$0.0785808  \\
15 & $~$0.082085 & -134.29 & -0.00909528 & $~$0.0608101 & $~$14923 & $~$9.05968 & -1220.85 & $~$8.54526 & $~$1.2767 & -0.0785808  \\
16 & $~$0.082085 & $~$134.29 & 	-0.00909528 & -0.0608101 & $~$14923 & -9.05968 & $~$1220.85 & $~$8.54526 & -1.2767 & $~$0.0785808  \\
\hline
17 & -0.110803 & -3.32012 & $~$0.0608101 & $~$13.4637 & $~$14906.6 & -491.394 & $~$1642.55 & $~$8.73603 & -0.0392728 & $~$0.530264  \\
18 & -0.110803 & $~$3.32012 & $~$0.0608101 & -13.4637 & $~$14906.6 & $~$491.394 & 	-1642.55 & $~$8.73603 & $~$0.0392728 & -0.530264  \\
19 & $~$0.110803 & -3.32012 & -0.0608101 & $~$13.4637 & $~$14906.6 & $~$491.394 & 	-1642.55 & $~$8.73603 & $~$0.0392728 & -0.530264  \\
20 & $~$0.110803 & $~$3.32012 & -0.0608101 & -13.4637 & $~$14906.6 & -491.394 & $~$1642.55 & $~$8.73603 & -0.0392728 & $~$0.530264  \\
\hline
21 & -0.0608101 & $~$13.4637 & $~$0.110803 & -3.32012 & $~$14922.7 & $~$67.1443 & -905.787 & $~$8.72662 & $~$0.287418 & -0.961579  \\
22 & -0.0608101 & -13.4637 & $~$0.110803 & $~$3.32012 & $~$14922.7 & -67.1443 & $~$905.787 & $~$8.72662 & -0.287418 & $~$0.961579 \\
23 & $~$0.0608101 & $~$13.4637 & -0.110803 & -3.32012 & $~$14922.7 & -67.1443 & $~$905.787 & $~$8.72662 & -0.287418 & $~$0.961579  \\
24 & $~$0.0608101 & -13.4637 & -0.110803 & $~$3.32012 & $~$14922.7 & $~$67.1443 & -905.787 & $~$8.72662 & $~$0.287418 & -0.961579 \\
\hline
25 & $~$0.0247235 & -40.4473 & -0.149569 & $~$14.8797 & $~$14923 & $~$9.11516 & -368.836 & $~$8.73535 & $~$0.0858003 & -1.29222  \\
26 & $~$0.0247235 & $~$40.4473 & -0.149569 & -14.8797 & $~$14923 & -9.11516 & $~$368.836 & $~$8.73535 & -0.0858003 & $~$1.29222 \\
27 & -0.0247235 & $~$40.4473 & $~$0.149569 & -14.8797 & $~$14923 & $~$9.11516 & -368.836 & $~$8.73535 & $~$0.0858003 & -1.29222 \\
28 & -0.0247235 & -40.4473 & $~$0.149569 & $~$14.8797 & $~$14923 & -9.11516 & $~$368.836 & $~$8.73535 & -0.0858003 & $~$1.29222  \\
\hline
29 & $~$0.149569 & -14.8797 & -0.0247235 & $~$40.4473 & 	14921.5 & 	$~$146.962 & 	-2207.36 & 	8.73621 & 	$~$0.00533148 & 	-0.215923 \\
30 & $~$0.149569 & $~$14.8797 & -0.0247235 & -40.4473 & 	14921.5 & 	-146.962 & 	$~$2207.36 & 	8.73621 & 	-0.00533148 & 	$~$0.215923   \\
31 & -0.149569 & -14.8797 & $~$0.0247235 & $~$40.4473 & 	14921.5 & 	-146.962 & 	$~$2207.36 & 	8.73621 & 	-0.00533148 & 	$~$0.215923  \\
32 & -0.149569 & $~$14.8797 & $~$0.0247235 & -40.4473 & 	14921.5 & 	$~$146.962 & 	-2207.36 & 	8.73621 & 	$~$0.00533148 & 	-0.215923  \\
\hline
\end{tabular}
\caption{\label{tab:i}
According to Equations (34) to (37) in \cite{Lin2021}, there are 32 candidate sets of ten parameters that yield the same predictions for the CKM elements. These sets satisfy the following values: $\vert V_{ud} \vert = \vert V_{tb} \vert \approx 0.9925$, $\vert V_{ub} \vert = \vert V_{td} \vert \approx 0.0075$, $\vert V_{us} \vert = \vert V_{ts} \vert = \vert V_{cd} \vert = \vert V_{cb} \vert \approx 0.122023$, and $\vert V_{cs} \vert \approx 0.9845$.
 }
\end{table*}

\end{document}